\documentclass{desyproc}
\usepackage{amsmath}
\numberwithin{equation}{section}
\newcommand{\MS}{\ensuremath{\overline{\text{MS}}}}

\newcommand{\1}[1]{\mathbf{1}^{#1}}
\newcommand{\2}[1]{\mathbf{2}^{#1}}
\newcommand{\3}[1]{\mathbf{3}^{#1}}
\newcommand{\4}[1]{\mathbf{4}^{#1}}
\newcommand{\5}[1]{\mathbf{5}^{#1}}

\begin{document}
\title{Higher radiative corrections in HQET}

\author{{\slshape Andrey Grozin}\\[1ex]
Budker Institute of Nuclear Physics, Novosibirsk 630090, Russia}

\contribID{grozin\_andrey}

\desyproc{DESY-PROC-2008-xx}
\acronym{HQP08} 
\doi  

\maketitle

\begin{abstract}
After a brief introduction to Heavy Quark Effective Theory,
we discuss $\alpha$ representation in HQET
and methods of calculation of some kinds of HQET diagrams
up to three loops.
\end{abstract}

\section{Introduction}
\label{S:Intro}

Effective field theories are very useful for describing physics
at low energies $\ll M$, or large distances $\gg1/M$,
where $M$ is a high energy scale where some new particles
or interactions become important.
Effective Lagrangians are constructed as series in $1/M$.
Coefficients in them are obtained by matching scattering amplitudes
in the full theory and in the effective one up to some order in $1/M$.
These matching coefficients are the only quantities which depend on $M$.
All calculations inside the effective theory involve only
characteristic energy scales (they are $\ll M$)
of processes under consideration.
The case when there is one such scale is especially simple.
We can choose the renormalization scale $\mu$ of order of
this characteristic energy scale.
Then there will be no large logarithms in perturbative series,
and truncating such series will produce small errors.
If we try to consider the same process in the full theory,
there is a second scale $M$,
and no choice of $\mu$ allows us to get rid of large logarithms.
Also, each extra scale in Feynman diagrams (with loops)
makes their calculation much more difficult technically.

Heavy Quark Effective Theory (HQET) is an effective low-energy
field theory for some problems in QCD.
In many widely-known effective field theories
(Heisenberg--Euler theory of low-energy photon interactions,
Fermi 4-fermion theory of weak interactions at low energies)
the heavy particle (electron or $W$ in these examples)
does not appear.
In HQET, the heavy quark appears in initial and final states,
but is always nearly on-shell and non-relativistic
(in some reference frame).

HQET is discussed in textbooks~\cite{MW:00,G:04} in detail.
Here we shall concentrate on methods and results
of calculations of multiloop Feynman diagrams in HQET,
see e.g. \cite{G:07}.
Various methods of multiloop calculations are presented
in the excellent book~\cite{S:06} in great detail,
most of these methods are used in HQET.

\section{Heavy Quark Effective Theory}
\label{S:HQET}

\subsection{Lagrangian and Feynman rules}
\label{S:Lagr}

We are going to consider a class of QCD problems involving
a single heavy quark with (on-shell) mass $m\gg\Lambda_{\text{QCD}}$.
Namely, we require that there exists a reference frame
where it stays nearly at rest all the time.
In other words, there exists a 4-velocity $v$ ($v^2=1$)
such that
\begin{equation}
p = mv + \tilde{p}\,,
\label{Lagr:p}
\end{equation}
and the characteristic residual momentum $\tilde{p}^\mu \ll m$.
Light quarks and gluons also have characteristic momenta
$p_i^\mu \ll m$.
Such problems can be described, instead of QCD,
by a simpler effective field theory called HQET.
Its Lagrangian is a series in $1/m$.
At the leading order,
\begin{equation}
L = \overline{\tilde{Q}}_v i v \cdot D \tilde{Q}_v
+ \mathcal{O}\left(\frac{1}{m}\right)
+ (\text{light fields})\,.
\label{Lagr:L}
\end{equation}
The HQET heavy-quark field satisfies
$\rlap/v \tilde{Q}_v = \tilde{Q}_v$.
All light fields are described as in QCD.
In the $v$ rest frame,
\begin{equation}
L = \tilde{Q}^+ i D_0 \tilde{Q}
+ \mathcal{O}\left(\frac{1}{m}\right)
+ (\text{light fields})\,,
\label{Lagr:L0}
\end{equation}
where $\tilde{Q}$ is a 2-component spinor.

The mass shell of the heavy quark,
i.e., the dependence of its residual energy $\tilde{p}_0$
on its momentum $\vec{p}$, is
\begin{equation*}
\tilde{p}_0 = p_0 - m = \frac{\vec{p}\,^2}{2m}\,.
\end{equation*}
At the leading order in $1/m$, it becomes $\tilde{p}_0 = 0$.
This is exactly what follows from the Lagrangian~(\ref{Lagr:L0}).

The HQET Lagrangian~(\ref{Lagr:L}) is not Lorentz-invariant,
because it contains a fixed vector $v$.
However, $v$ is not uniquely defined.
It can be changed by $\sim\tilde{p}/m$ (see~(\ref{Lagr:p})).
Lagrangians with such different choices of $v$
must produce identical physical predictions.
This requirement is called reparametrization invariance,
and it restricts $1/m^n$ corrections in the Lagrangian.

The heavy-quark chromomagnetic moment is,
by dimensionality, $\sim1/m$.
Therefore, at the leading order
the heavy-quark spin does not interact with the gluon field.
We may rotate the spin at will without changing physics
--- heavy-quark spin symmetry.
In particular, $B$ and $B^*$ are degenerate
and have identical properties,
because they can be transformed into each other
by rotating the $b$ spin.
We can even change the magnitude of the heavy-quark spin
(e.g., to switch it off) without changing physics
--- this supersymmetry group is called superflavour symmetry.

It is difficult to simulate a QCD heavy quark on the lattice
because the lattice spacing $a$ must be much less than
the minimum characteristic distance of the problem, $1/m$.
The HQET Lagrangian does not contain $m$,
and the only applicability condition of its discretization
is $a\ll1/\tilde{p}$.
When we investigate the structure of heavy--light hadrons,
$\tilde{p}\sim\Lambda_{\text{QCD}}$,
and the condition $a\ll1/\Lambda_{\text{QCD}}$
is the same as for light hadrons.

The Lagrangian~(\ref{Lagr:L}) gives the Feynman rules
\begin{equation}
\raisebox{-3.5mm}{\begin{picture}(22,6.5)
\put(11,4.5){\makebox(0,0){\includegraphics{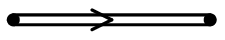}}}
\put(11,0){\makebox(0,0)[b]{$\tilde{p}$}}
\end{picture}} =
i\,\frac{1+\rlap/v}{2}\,
\frac{1}{\tilde{p}\cdot v+i0}\,,\qquad
\raisebox{-1mm}{\begin{picture}(22,15)
\put(11,6.5){\makebox(0,0){\includegraphics{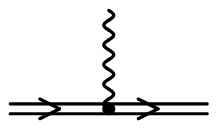}}}
\put(9,15){\makebox(0,0)[t]{$a$}}
\put(13,15){\makebox(0,0)[t]{$\mu$}}
\end{picture}} =
i g t^a v^\mu\,.
\label{Feyn:rules}
\end{equation}
In the $v$ rest frame, the propagator is
(the unit $2\times2$ spin matrix assumed)
\begin{equation}
i\,\frac{1}{\tilde{p}_0+i0}\,.
\label{Feyn:v0}
\end{equation}
In the coordinate space, the heavy quark does not move:
\begin{equation}
\raisebox{-3.5mm}{\begin{picture}(22,6.5)
\put(11,4.5){\makebox(0,0){\includegraphics{grozin_andrey.fig01.eps}}}
\put(1,0){\makebox(0,0)[b]{$0$}}
\put(21,0){\makebox(0,0)[b]{$x$}}
\end{picture}} =
-i \theta(x_0) \delta(\vec{x}\,)\,.
\label{Feyn:coord}
\end{equation}
HQET-quark loops vanish because the heavy quark propagates
only forward in time.
We can also see this in the momentum space:
all poles of the propagators in such a loop
are in the lower $\tilde{p}_0$ half-plane,
and closing the integration contour upwards, we get 0.

These Feynman rules can be also obtained from QCD at $m\to\infty$.
The QCD massive-quark propagator gives the HQET one:
\begin{align}
\raisebox{-3.5mm}{\begin{picture}(22,6.5)
\put(11,4.5){\makebox(0,0){\includegraphics{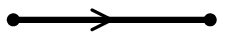}}}
\put(11,0){\makebox(0,0)[b]{$mv+\tilde{p}$}}
\end{picture}} &=
\raisebox{-3.5mm}{\begin{picture}(22,6.5)
\put(11,4.5){\makebox(0,0){\includegraphics{grozin_andrey.fig01.eps}}}
\put(11,0){\makebox(0,0)[b]{$\tilde{p}$}}
\end{picture}}
+ \mathcal{O}\left(\frac{\tilde{p}}{m}\right)\,,
\nonumber\\
\frac{m + m\rlap/v + \rlap/\tilde{p}}{(mv+\tilde{p})^2-m^2+i0} &=
\frac{1+\rlap/v}{2}\,
\frac{1}{\tilde{p}\cdot v+i0}
+ \mathcal{O}\left(\frac{\tilde{p}}{m}\right)\,.
\label{Feyn:QCD}
\end{align}
The QCD vertex, when sandwiched between two projectors,
becomes the HQET one:
\begin{equation}
\frac{1+\rlap/v}{2} \gamma^\mu \frac{1+\rlap/v}{2} =
\frac{1+\rlap/v}{2} v^\mu \frac{1+\rlap/v}{2}\,.
\label{Feyn:vertex}
\end{equation}
When there is an external leg near a vertex,
there is no projector;
but we can insert it, and the argument holds.

We have thus proved that at the tree level any QCD diagram
is equal to the corresponding HQET diagram up to
$\mathcal{O}(\tilde{p}/m)$ corrections.
This is not true at loops,
because loop momenta can be arbitrarily large.
Renormalization properties of HQET
(anomalous dimensions, etc.) differ from those in QCD.

\subsection{One-loop propagator diagram}
\label{S:L1}

\begin{figure}[ht]
\begin{center}
\begin{picture}(54,24)
\put(27,12.5){\makebox(0,0){\includegraphics{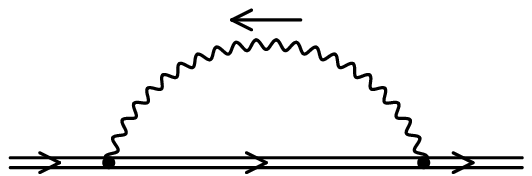}}}
\put(27,0){\makebox(0,0)[b]{$k+\tilde{p}$}}
\put(27,24){\makebox(0,0)[t]{$k$}}
\put(6,0){\makebox(0,0)[b]{$\tilde{p}$}}
\put(48,0){\makebox(0,0)[b]{$\tilde{p}$}}
\put(27,9){\makebox(0,0)[t]{$n_1$}}
\put(27,13){\makebox(0,0)[b]{$n_2$}}
\end{picture}
\end{center}
\caption{One-loop propagator diagram}
\label{F:h1}
\end{figure}

Let's calculate the simplest one-loop diagram (Fig.~\ref{F:h1})
\begin{align}
&\frac{1}{i\pi^{d/2}} \int \frac{d^d k}%
{\bigl[-2\left(k+\tilde{p}\right)\cdot v-i0\bigr]^{n_1}
\bigl[-k^2-i0\bigr]^{n_2}} =
\nonumber\\
&\frac{1}{i\pi^{d/2}} \int \frac{d k_0\,d^{d-1} \vec{k}}%
{\bigl[-2\left(k_0+\omega\right)-i0\bigr]^{n_1}
\bigl[-k_0^2+\vec{k}\,^2-i0\bigr]^{n_2}}
= (-2\omega)^{d-n_1-2n_2} I(n_1,n_2)\,.
\label{L1:def}
\end{align}
It depends only on the residual energy $\omega=\tilde{p}_0$,
not $\vec{\tilde{p}}$;
the power of $-2\omega$ is clear from dimensional counting.

If $\omega>0$, real pair production is possible,
and we are on a cut.
We shall consider the case $\omega<0$,
when the integral is an analytic function of $\omega$.
We'll set $-2\omega=1$.
If $n_1$ is integer and $n_1\le0$,
$I(n_1,n_2)=0$ because this is a massless vacuum diagram.
If $n_2$ is integer and $n_2\le0$,
$I(n_1,n_2)=0$ because the diagram contains an HQET loop.

\begin{figure}[ht]
\begin{center}
\begin{picture}(42,42)
\put(21,21){\makebox(0,0){\includegraphics{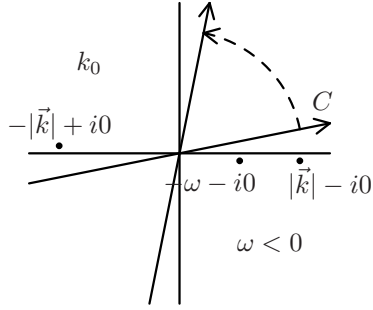}}}
\put(40,28){\makebox(0,0){$C$}}
\put(41,17){\makebox(0,0){$|\vec{k}|-i0$}}
\put(5,25){\makebox(0,0){$-|\vec{k}|+i0$}}
\put(25,17){\makebox(0,0){$-\omega-i0$}}
\put(33,9){\makebox(0,0){$\omega<0$}}
\put(9,33){\makebox(0,0){$k_0$}}
\end{picture}
\end{center}
\caption{Wick rotation}
\label{F:Wick}
\end{figure}

At $\omega<0$, all poles in the $k_0$ plane are
below the real axis at $k_0>0$
and above the real axis at $k_0<0$,
and we can rotate the integration contour counterclockwise 
without crossing poles
(if $\omega>0$, we cross the pole at $k_0=-\omega-i0$).
This Wick rotation
\begin{equation}
k_0 = i k_{E0}
\label{L1:Wick}
\end{equation}
(Fig.~\ref{F:Wick})
brings us into Euclidean momentum space ($k^2=-k_E^2$).

\begin{figure}[ht]
\begin{center}
\begin{picture}(34,34)
\put(17,17){\makebox(0,0){\includegraphics{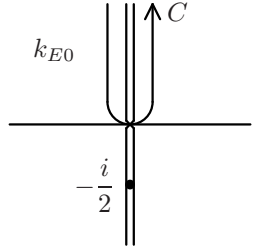}}}
\put(22,32){\makebox(0,0)[l]{$C$}}
\put(15,9){\makebox(0,0)[r]{$\displaystyle
- \frac{i}{2}$}}
\put(7,27){\makebox(0,0){$k_{E0}$}}
\end{picture}
\end{center}
\caption{Integration contour}
\label{F:Contour0}
\end{figure}

In the Euclidean space,
\begin{equation}
I(n_1,n_2) = \frac{1}{\pi^{d/2}}
\int \frac{d k_{E0}}{\bigl(1 - 2 i k_{E0}\bigr)^{n_1}}
\int \frac{d^{d-1} \vec{k}}{\bigl(\vec{k}\,^2+k_{E0}^2\bigr)^{n_2}}\,.
\label{L1:Eucl}
\end{equation}
Using the well-known formula
\begin{equation}
\int \frac{d^d k_E}{(k_E^2+m^2)^n} = \pi^{d/2} (m^2)^{d/2-n}
\frac{\Gamma\bigl(n-\frac{d}{2}\bigr)}{\Gamma(n)}
\label{L1:Tadpole}
\end{equation}
with $d\to d-1$, $m^2\to k_{E0}^2$, $n\to n_2$, we obtain
\begin{equation}
I(n_1,n_2) =
\frac{\Gamma\bigl(n_2-\frac{d-1}{2}\bigr)}{\pi^{1/2} \Gamma(n_2)}
\int \frac{\bigl(k_{E0}^2\bigr)^{(d-1)/2-n_2} d k_{E0}}%
{\bigl(1 - 2 i k_{E0}\bigr)^{n_1}}\,.
\label{L1:kE0}
\end{equation}
The integrand is even in $k_{E0}$, and has cuts at $k_{E0}^2<0$
(it would be wrong to write it as $k_{E0}^{d-1-2n_2}$).
Deforming the integration contour around the upper cut
(Fig.~\ref{F:Contour0}),
we can express the integral via the discontinuity
at this cut:
\begin{equation}
I(n_1,n_2) = 2
\frac{\Gamma\bigl(n_2-\frac{d-1}{2}\bigr)}{\pi^{1/2} \Gamma(n_2)}
\cos\left[\pi\left(\frac{d}{2}-n_2\right)\right]
\int_0^\infty \frac{k^{d-1-2n_2} d k}{(2k+1)^{n_1}}\,.
\label{L1:cut}
\end{equation}

This integral can be easily calculated in $\Gamma$ functions:
\begin{equation}
I(n_1,n_2) = \frac{2^{2n_2-d+1}}{\pi^{1/2}}
\cos\left[\pi\left(\frac{d}{2}-n_2\right)\right]
\frac{\Gamma(d-2n_2) \Gamma(n_1+2n_2-d)
\Gamma\bigl(n_2-\frac{d-1}{2}\bigr)}%
{\Gamma(n_1) \Gamma(n_2)}\,.
\label{L1:I0}
\end{equation}
Using the well-known properties of the $\Gamma$ function
\begin{equation}
\Gamma(2x) = \pi^{-1/2} 2^{2x-1} \Gamma(x) \Gamma\bigl(x+\tfrac{1}{2}\bigr)\,,
\qquad
\Gamma(x) \Gamma(1-x) = \frac{\pi}{\sin\pi x}\,,
\label{L1:Gamma}
\end{equation}
we can simplify this result:
\begin{equation}
I(n_1,n_2) =
\frac{\Gamma(n_1+2n_2-d) \Gamma\bigl(\frac{d}{2}-n_2\bigr)}%
{\Gamma(n_1) \Gamma(n_2)}\,.
\label{L1:I}
\end{equation}

\begin{figure}[t]
\begin{center}
\begin{picture}(32,15.5)
\put(16,7.75){\makebox(0,0){\includegraphics{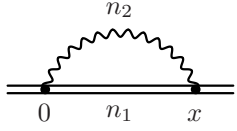}}}
\put(16,0){\makebox(0,0)[b]{$n_1$}}
\put(16,15.75){\makebox(0,0)[t]{$n_2$}}
\put(6,0){\makebox(0,0)[b]{$0$}}
\put(26,0){\makebox(0,0)[b]{$x$}}
\end{picture}
\end{center}
\caption{One-loop propagator diagram in coordinate space}
\label{F:Coord}
\end{figure}

It is also easy to derive this result in coordinate space~\cite{G:07}.
HQET propagators in momentum and coordinate space are related by
\begin{align}
&\int_{-\infty}^{+\infty} \frac{e^{-i\omega t}}{(-2\omega-i0)^n}
\frac{d\omega}{2\pi} =
\frac{i}{2\Gamma(n)} \left(\frac{it}{2}\right)^{n-1}
e^{-0t} \theta(t)\,,
\label{L1:Fh1}\\
&\int_0^\infty e^{(i\omega-0)t} \left(\frac{it}{2}\right)^{n-1} dt =
- \frac{2 i \Gamma(n)}{(-2\omega-i0)^n}\,;
\label{L1:Fh2}
\end{align}
massless propagators --- by
\begin{align}
&\int \frac{e^{-ip\cdot x}}{(-p^2-i0)^n} \frac{d^d p}{(2\pi)^d} =
\frac{i}{(4\pi)^{d/2}} \frac{\Gamma(d/2-n)}{\Gamma(n)}
\left(\frac{4}{-x^2+i0}\right)^{d/2-n}\,,
\label{L1:Fl1}\\
&\int \left(\frac{4}{-x^2+i0}\right)^n e^{ip\cdot x} d^d x =
- i (4\pi)^{d/2} \frac{\Gamma(d/2-n)}{\Gamma(n)}
\frac{1}{(-p^2-i0)^{d/2-n}}\,.
\label{L1:Fl2}
\end{align}
Our diagram in coordinate space (Fig.~\ref{F:Coord}, $x=vt$)
is just the product of the heavy propagator~(\ref{L1:Fh1})
and the light one~(\ref{L1:Fl1}) (where $-x^2/4=-t^2/4=(it/2)^2$):
\begin{equation*}
- \frac{1}{2} \frac{1}{(4\pi)^{d/2}}
\frac{\Gamma(d/2-n_2)}{\Gamma(n_1) \Gamma(n_2)}
\left(\frac{it}{2}\right)^{n_1+2n_2-d-1} \theta(t)\,.
\end{equation*}
The inverse Fourier transform~(\ref{L1:Fh2})
gives our diagram~(\ref{L1:def}) in momentum space
\begin{equation*}
\frac{i}{(4\pi)^{d/2}} I(n_1,n_2) (-2\omega)^{d-n_1-2n_2}\,,
\end{equation*}
where $I(n_1,n_2)$ is given by~(\ref{L1:I}).

\subsection{Renormalization}
\label{S:Ren}

The Lagrangian contains bare fields and parameters:
\begin{equation}
L = \overline{\tilde{Q}}_{v0} i v \cdot D_0 \tilde{Q}_{v0}\qquad
D_{0\mu} = \partial_\mu - i g_0 A^a_{0\mu} t^a\,.
\label{Ren:L}
\end{equation}
They are related to the renormalized ones
by the renormalization constants:
\begin{equation}
\tilde{Q}_{v0} = \tilde{Z}_Q^{1/2} \tilde{Q}_v\,,\quad
A_0 = Z_A^{1/2} A\,,\quad
a_0 = Z_A a\,,\quad
g_0 = Z_\alpha^{1/2} g\,,
\label{Ren:Ren}
\end{equation}
where $a_0$ is the gauge-fixing parameter.
The ghost field, the light-quark fields (and their masses)
are renormalized as in QCD (not written here).
Minimal renormalization constants have the structure
\begin{equation}
Z_i = 1 + \frac{Z_{11}}{\varepsilon} \frac{\alpha_s}{4\pi}
+ \left(\frac{Z_{22}}{\varepsilon^2} + \frac{Z_{21}}{\varepsilon}\right)
\left(\frac{\alpha_s}{4\pi}\right)^2
+ \cdots
\label{Ren:Min}
\end{equation}
They don't contain $\varepsilon^0$ and $\varepsilon^n$ ($n>0$) terms,
only negative powers needed to remove divergences,
and hence are called minimal.
We have to define $\alpha_s$ to be exactly dimensionless.
In the \MS{} scheme $\alpha_s$ depends on the renormalization scale $\mu$:
\begin{equation}
\frac{g_0^2}{(4\pi)^{d/2}} = \mu^{2\varepsilon} \frac{\alpha_s(\mu)}{4\pi}
Z_\alpha(\alpha_s(\mu)) e^{\gamma_E \varepsilon}\,,
\label{Ren:MSbar}
\end{equation}
where $\gamma_E$ is the Euler constant.

\begin{figure}[ht]
\begin{center}
\begin{picture}(54,24)
\put(27,12.5){\makebox(0,0){\includegraphics{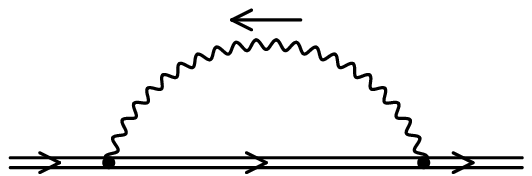}}}
\put(27,0){\makebox(0,0)[b]{$k+\tilde{p}$}}
\put(27,24){\makebox(0,0)[t]{$k$}}
\put(6,0){\makebox(0,0)[b]{$\tilde{p}$}}
\put(48,0){\makebox(0,0)[b]{$\tilde{p}$}}
\end{picture}
\end{center}
\caption{One-loop heavy-quark self-energy}
\label{F:se1}
\end{figure}

Let's calculate the HQET propagator with one-loop accuracy:
\begin{align}
&\raisebox{-1.25mm}{\includegraphics{grozin_andrey.fig01.eps}} +
\raisebox{-3.75mm}{\includegraphics{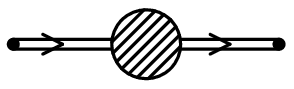}}
+ \cdots
\nonumber\\
&i \tilde{S}(\omega) = i \tilde{S}_0(\omega)
+ i \tilde{S}_0(\omega) (-i) \tilde{\Sigma}(\omega) i \tilde{S}_0(\omega)
+ \cdots
\label{Ren:Prop}
\end{align}
where
\begin{equation}
\tilde{S}_0(\omega) = \frac{1}{\omega}\,.
\label{Ren:S0}
\end{equation}
The one-loop heavy-quark self-energy (Fig.~\ref{F:se1}) is
\begin{equation}
\tilde{\Sigma}(\omega) = i C_F \int \frac{d^d k}{(2\pi)^d}
i g_0 v^\mu \frac{1}{k_0+\omega} i g_0 v^\nu
\frac{-i}{k^2} \left(g_{\mu\nu} - \xi \frac{k_\mu k_\nu}{k^2}\right)\,,
\label{Ren:se1}
\end{equation}
where $\xi=1-a_0$.
In the numerator, we may replace
$(k\cdot v)^2=(k_0+\omega-\omega)^2\to\omega^2$,
because if we cancel $k_0+\omega$ in the denominator
the integral vanishes.
Using~(\ref{L1:I}), we obtain
\begin{align}
\tilde{\Sigma}(\omega) &= C_F
\frac{g_0^2 (-2\omega)^{1-2\varepsilon}}{(4\pi)^{d/2}}
\left[2 I(1,1) + \frac{\xi}{2} I(1,2)\right]
\nonumber\\
&= C_F
\frac{g_0^2 (-2\omega)^{1-2\varepsilon}}{(4\pi)^{d/2}}
\frac{\Gamma(1+2\varepsilon) \Gamma(1-\varepsilon)}{d-4}
\left(\xi + \frac{2}{d-3}\right)\,.
\label{Ren:Sigma1}
\end{align}

The propagator expressed via renormalized quantities is
\begin{equation}
\omega \tilde{S}(\omega) = 1 + C_F \frac{\alpha_s(\mu)}{4\pi\varepsilon}
e^{-2L\varepsilon} \bigl[3 - a(\mu) + 4\varepsilon + \cdots\bigr]\,,
\label{Ren:S1}
\end{equation}
where
\begin{equation*}
L = \log\frac{-2\omega}{\mu}\,.
\end{equation*}
It should be equal
$\tilde{S}(\omega)=\tilde{Z}_Q \tilde{S}_r(\omega)$,
where the renormalized propagator $\tilde{S}_r(\omega)$
is finite at $\varepsilon\to0$.
Therefore,
\begin{equation}
\tilde{Z}_Q = 1 + C_F (3-a) \frac{\alpha_s}{4\pi\varepsilon}
\label{Ren:ZQ}
\end{equation}
(it is also easy to write $\tilde{S}_r(\omega)$).

The HQET field does not renormalized in the Yennie gauge $a=3$.
This is exactly the reason why this gauge has been introduced
in the theory of electrons interacting with soft photons
(Bloch--Nordsieck model), which is the Abelian HQET.
In the Abelian case, this non-renormalization holds to all orders
due to the exponentiation theorem.
In HQET, this is only true at one loop.

\begin{figure}[ht]
\begin{center}
\begin{picture}(102,17)
\put(51,8.5){\makebox(0,0){\includegraphics{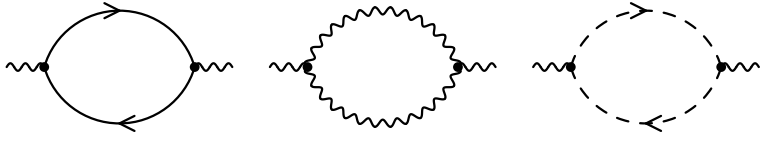}}}
\end{picture}
\end{center}
\caption{One-loop gluon self-energy}
\label{F:gluon}
\end{figure}

Now we shall discuss the renormalization of $g$.
Due to the gauge invariance, all $g$'s in the Lagrangian are equal.
The coupling of the HQET quark field to gluon
is thus identical to the usual QCD coupling,
where the HQET heavy flavour is not counted
in the number of flavours $n_f$.
In order to find $Z_\alpha$ we need to renormalize
the heavy-quark -- gluon vertex and all propagators
attached to it.
We have already calculated $\tilde{Z}_Q$.
The renormalization of the gluon propagator is well known
(Fig.~\ref{F:gluon}):
\begin{equation}
Z_A = 1 - \left[ \frac{C_A}{2} \left(a - \frac{13}{3}\right)
+ \frac{4}{3} T_F n_f \right] \frac{\alpha_s}{4\pi\varepsilon}\,.
\label{Ren:ZA}
\end{equation}

Let's introduce the vertex
\begin{equation}
\raisebox{-3.75mm}{\begin{picture}(25,15)
\put(12.5,8.75){\makebox(0,0){\includegraphics{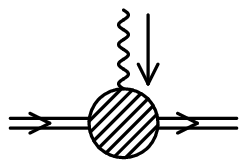}}}
\put(5,0){\makebox(0,0)[b]{$\omega$}}
\put(20,0){\makebox(0,0)[b]{$\omega'$}}
\put(18,10){\makebox(0,0)[l]{$q$}}
\end{picture}} = i g_0 t^a \tilde{\Gamma}^\mu\,,\quad
\tilde{\Gamma}^\mu = v^\mu + \tilde{\Lambda}^\mu\,,
\label{Ren:Vdef}
\end{equation}
where $\tilde{\Lambda}^\mu$ starts from one loop.
When expressed via renormalized quantities,
the vertex should be
$\tilde{\Gamma}=\tilde{Z}_\Gamma\tilde{\Gamma}_r$,
where the renormalized vertex $\tilde{\Gamma}_r$
is finite at $\varepsilon\to0$.
A physical matrix element is obtained from the corresponding vertex
by multiplying it by the wave-function renormalization constant
$Z_i^{1/2}$ for each external leg.
In our case,
\begin{equation*}
g_0 \tilde{\Gamma} \tilde{Z}_Q Z_A^{1/2}
= g \tilde{\Gamma}_r Z_\alpha^{1/2} \tilde{Z}_\Gamma \tilde{Z}_Q Z_A^{1/2}
= \text{finite}.
\end{equation*}
Therefore, $Z_\alpha^{1/2} \tilde{Z}_\Gamma \tilde{Z}_Q Z_A^{1/2}$
must be finite.
But the only minimal~(\ref{Ren:Min}) renormalization constant
finite at $\varepsilon\to0$ is 1:
\begin{equation}
Z_\alpha = \bigl(\tilde{Z}_\Gamma \tilde{Z}_Q\bigr)^{-2} Z_A^{-1}\,.
\label{Ren:Za}
\end{equation}

\begin{figure}[b]
\begin{center}
\begin{picture}(65,16)
\put(15,8){\makebox(0,0){\includegraphics{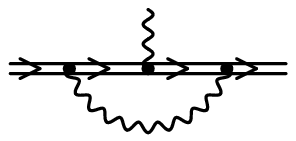}}}
\put(50,8){\makebox(0,0){\includegraphics{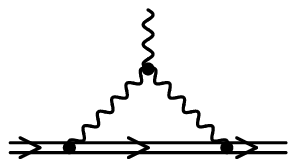}}}
\end{picture}
\end{center}
\caption{One-loop HQET vertex}
\label{F:Lambda1}
\end{figure}

At one loop the HQET vertex is given by two diagrams
(Fig.~\ref{F:Lambda1}).
It is very easy to calculate the first one.
It contains two heavy denominators which can be replaced
by a difference:
\begin{equation}
\frac{1}{(k_0+\omega)(k_0+\omega')} = \frac{1}{\omega'-\omega}
\left(\frac{1}{k_0+\omega} - \frac{1}{k_0+\omega'}\right)\,.
\label{Ren:parfrac}
\end{equation}
We get the difference of the self-energies:
\begin{equation}
\tilde{\Lambda}_1^\mu = - \left(1 - \frac{C_A}{2 C_F}\right)
\frac{\tilde{\Sigma}(\omega')-\tilde{\Sigma}(\omega)}{\omega'-\omega}
v^\mu\,.
\label{Ren:Lambda1}
\end{equation}
This result can also be obtained from the Ward identity.
The UV divergence of this contribution is
\begin{equation}
\tilde{\Lambda}_1^\mu = \left(C_F - \frac{C_A}{2}\right)
(a-3) \frac{\alpha_s}{4\pi\varepsilon} v^\mu\,.
\label{Ren:UV1}
\end{equation}

The second diagram is more difficult.
It has been calculated in~\cite{DG:01}.
Now we only need its UV divergence,
and it should be $\sim v^\mu$.
We may nullify all external momenta.
After that, the diagram will contain no scale and hence vanish.
It will contain both UV and IR divergences which cancel.
Therefore, we'll have to introduce some IR regularization
to get the UV divergence.
We have
\begin{align}
i g_0 \tilde{\Lambda}_2^\mu =& \frac{C_A}{2}
\int \frac{d^d k}{(2\pi)^d}
i g_0 v^{\alpha'} \frac{i}{k\cdot v} i g_0 v^{\beta'}
\nonumber\\
&{}\times\frac{-i}{k^2}
\left( g_{\alpha\alpha'} - \xi \frac{k_\alpha k_{\alpha'}}{k^2} \right)
\frac{-i}{k^2}
\left( g_{\beta\beta'} - \xi \frac{k_\beta k_{\beta'}}{k^2} \right)
i g_0 V^{\alpha\beta\mu}(k,-k,0)\,,
\label{Ren:V2a}
\end{align}
where the three-gluon vertex is
\begin{equation}
V^{\mu_1\mu_2\mu_3}(p_1,p_2,p_3)
= (p_3-p_2)^{\mu_1} g^{\mu_2\mu_3}
+ (p_1-p_3)^{\mu_2} g^{\mu_3\mu_1}
+ (p_2-p_1)^{\mu_3} g^{\mu_1\mu_2}\,.
\label{Ren:g3}
\end{equation}
It vanishes when contracted with the same vector in all three indices:
\begin{equation*}
V^{\mu_1\mu_2\mu_3}(p_1,p_2,p_3) v_{\mu_1} v_{\mu_2} v_{\mu_3} = 0\,,
\end{equation*}
and when contracted in two indices with the corresponding momenta:
\begin{equation*}
V^{\mu_1\mu_2\mu_3}(p_1,p_2,p_3) p_{1\,\mu_1} p_{2\,\mu_2} = 0\,.
\end{equation*}
Therefore, $\xi^0$ and $\xi^2$ terms vanish:
\begin{equation*}
\tilde{\Lambda}_2^\mu v_\mu = i C_A g_0^2 \xi
\int \frac{d^d k}{(2\pi)^d}
\frac{k^2-(k\cdot v)^2}{(k^2)^3}\,.
\end{equation*}
Averaging over $k$ directions $(k\cdot v)^2\to k^2/d$,
we get
\begin{equation*}
\tilde{\Lambda}_2^\mu v_\mu = i C_A g_0^2 \xi
\left(1 - \frac{1}{d}\right)
\int \frac{d^d k}{(2\pi)^d} \frac{1}{(k^2)^2}\,.
\end{equation*}
The UV divergence of this integral can be obtained by introducing
any IR regularization, e.g.,
an IR cut-off in the Euclidean momentum integral
or a small mass:
\begin{equation}
\left. \int \frac{d^d k}{(2\pi)^d} \frac{1}{(k^2)^2} \right|_{\text{UV}}
= \frac{i}{(4\pi)^{d/2} \varepsilon}\,.
\label{Ren:UV}
\end{equation}
We arrive at the UV divergence of the second vertex diagram:
\begin{equation}
\tilde{\Lambda}_2^\mu = - \frac{3}{4} C_A (1-a)
\frac{\alpha_s}{4\pi\varepsilon} v^\mu\,.
\label{Ren:UV2}
\end{equation}

From~(\ref{Ren:UV1}) and~(\ref{Ren:UV2}) we obtain
\begin{equation}
\tilde{Z}_\Gamma = 1 + \left[ C_F (a-3) + C_A \frac{a+3}{4} \right]
\frac{\alpha_s}{4\pi\varepsilon}\,.
\label{Ren:ZGamma}
\end{equation}
The product which appears in~(\ref{Ren:Za}) is
\begin{equation}
\tilde{Z}_\Gamma \tilde{Z}_Q = 1 + C_A \frac{a+3}{4}
\frac{\alpha_s}{4\pi\varepsilon}\,.
\label{Ren:ZGA}
\end{equation}
In the Abelian case $\tilde{Z}_\Gamma \tilde{Z}_Q=1$
to all orders, due to the Ward identity.
This is why we only got the non-abelian colour structure $C_A$
in~(\ref{Ren:ZGA}).
Finally, combining this with~(\ref{Ren:ZA}),
we see that the $a$ dependence cancels, and
\begin{equation}
Z_\alpha = 1 - \beta_0 \frac{\alpha_s}{4\pi\varepsilon}\,,\quad
\beta_0 = \frac{11}{3} C_A - \frac{4}{3} T_F n_f\,.
\label{Ren:Zalpha}
\end{equation}
Thus we have derived the one-loop $\beta$ function of QCD
($n_f$ does not include the HQET heavy flavour $Q$).
Most textbooks use the massless-quark -- gluon vertex
or the ghost -- gluon one (see, e.g., \cite{G:07}).
In the later case, calculations are a little shorter.
The HQET derivation presented here is as short as the ghost one.

\section{$\alpha$ parametrization}
\label{S:alpha}

\subsection{General formulae}
\label{S:ag}

$\alpha$ parametrization of Feynman integrals
(including those containing numerators)
is discussed in many textbooks, see e.g.~\cite{Z:79}.
Here we shall discuss HQET integrals;
all rules can be trivially obtained from~\cite{Z:79},
though they were not yet explicitly stated in the literature.

First let's calculate the one-loop diagram~(\ref{L1:def})
(Fig.~\ref{F:h1}) using $\alpha$ parametrization
\begin{equation}
\frac{1}{a^n} = \frac{1}{\Gamma(n)}
\int_0^\infty d\alpha\,\alpha^{n-1} e^{-a\alpha}\,.
\label{alpha:alpha}
\end{equation}
We get
\begin{equation*}
\frac{1}{\Gamma(n_1) \Gamma(n_2)}
\int d\alpha\,\alpha^{n_2-1}\,d\beta\,\beta^{n_1-1}\,d^d k\,e^X\,,\quad
X = \alpha k^2 + 2 \beta (k+\tilde{p})\cdot v\,.
\end{equation*}
We shift the integration momentum
\begin{equation*}
k = k' - \frac{\beta}{\alpha} v
\end{equation*}
to eliminate the linear term in the exponent:
\begin{equation*}
X = \alpha k^{\prime2} - \frac{\beta^2}{\alpha} + 2 \beta \omega\,.
\end{equation*}
Now it is easy to calculate the momentum integral:
\begin{equation}
\int d^d k\,e^{\alpha k^2}
= i \int d^d k_E\,e^{-\alpha k_E^2}
= i \left(\frac{\pi}{\alpha}\right)^{d/2}\,.
\end{equation}
Therefore,
\begin{equation*}
(-2\omega)^{d-n_1-2n_2} I(n_1,n_2) =
\frac{1}{\Gamma(n_1) \Gamma(n_2)}
\int d\alpha\,\alpha^{n_2-1}\,d\beta\,\beta^{n_1-1}\,\alpha^{-d/2}
\exp\left(-\frac{\beta^2}{\alpha}+2\beta\omega\right)\,.
\end{equation*}
Now we make the substitution $\beta=\alpha y$ and integrate in $\alpha$:
\begin{equation*}
\frac{\Gamma\bigl(n_1+n_2-\frac{d}{2}\bigr)}{\Gamma(n_1) \Gamma(n_2)}
\int_0^\infty d y\,y^{n_1-1}
\bigl[y(y-2\omega)\bigr]^{d/2-n_1-n_2}\,.
\end{equation*}
The HQET Feynman parameter $y$ has the dimensionality of energy
and varies from 0 to $\infty$.
The $y$ integral can be easily calculated in $\Gamma$ functions,
and we again obtain~(\ref{L1:I}).

Now we shall consider the most general HQET Feynman integral
without numerators.
Any HQET diagram contains a single heavy line and has the form
\begin{equation}
I = \int \prod \frac{d^d k_i}{i\pi^{d/2}}\,
\frac{1}{\prod L_a^{n_a}\,\prod H_c^{n_c}}\,,
\label{alpha:form}
\end{equation}
where $k_i$ are loop momenta ($i,j\in[1,L]$),
\begin{equation*}
L_a = m_a^2 - q_a^2 - i0\,,\quad
H_c = - 2 q_c \cdot v - i0
\end{equation*}
are light denominators ($a,b\in[1,N_l]$)
and heavy ones ($c,d\in[1,N_h]$):
\begin{equation*}
q_a = \sum N_{ai} k_i + \sum N_{an} p_n\,,\quad
q_c = \sum N_{ci} k_i + \sum N_{cn} p_n\,,
\end{equation*}
$p_n$ are external momenta ($n,m\in[1,N_e-1]$),
and the coefficients $N$ express momenta of propagators
via the loop and external momenta
(these coefficients are equal to 0 or $\pm1$).
Using the $\alpha$ representation~(\ref{alpha:alpha}) for all lines,
we obtain
\begin{align}
&I = \frac{1}{\prod \Gamma(n_a)\,\prod \Gamma(n_c)}
\int \prod d\alpha_a\,\alpha_a^{n_a-1}\, \prod d\beta_c\,\beta_c^{n_c-1}
\prod \frac{d^d k_i}{i\pi^{d/2}}\,e^X\,,
\nonumber\\
&X = \sum \alpha_a (q_a^2-m_a^2) + 2 \sum \beta_c q_c \cdot v\,.
\label{alpha:I}
\end{align}
Dimensionalities of the parameters are
$\alpha_a\sim1/M^2$, $\beta_c\sim1/M$.
The exponent is
\begin{equation}
X = \sum M_{ij} k_i \cdot k_j - 2 \sum Q_i \cdot k_i + Y\,,
\label{alpha:X}
\end{equation}
where
\begin{align}
&M_{ij} = \sum \alpha_a N_{ai} N_{aj}\,,
\nonumber\\
&Q_i = - \sum \alpha_a N_{ai} N_{an} p_n - v \sum \beta_c N_{ci}\,,
\nonumber\\
&Y = \sum \alpha_a \left(\sum N_{an} p_n\right)^2
+ 2 \sum \beta_c N_{cn} p_n \cdot v
- \sum \alpha_a m_a^2\,.
\label{alpha:MQY}
\end{align}

Now we shift the loop momenta $k_i=k'_i+K_i$ to eliminate linear terms:
\begin{equation}
K_i = \sum M^{-1}_{ij} Q_j\,.
\label{alpha:K}
\end{equation}
Then
\begin{equation}
X = \sum M_{ij} k'_i \cdot k'_j
- \sum M^{-1}_{ij} Q_i \cdot Q_j
+ Y\,.
\label{alpha:X2}
\end{equation}
Performing the Wick rotation to Euclidean $k'_i$
and integration in the loop momenta, we obtain
\begin{align}
&I = \frac{1}{\prod \Gamma(n_a)\,\prod \Gamma(n_c)}
\int \prod d\alpha_a\,\alpha_a^{n_a-1}\, \prod d\beta_c\,\beta_c^{n_c-1}
\left[D(\alpha)\right]^{-d/2}
\nonumber\\
&\hphantom{I={}}\times \exp \left[
- \frac{A(\alpha)+A_1(\alpha,\beta)+A_2(\alpha,\beta)}{D(\alpha)}
- \sum \alpha_a m_a^2 \right]\,,
\label{alpha:res}
\end{align}
where
\begin{align}
&D(\alpha) = \det M\,,
\nonumber\\
&\frac{A(\alpha)}{D(\alpha)} =
\sum M^{-1}_{ij} \alpha_a \alpha_b N_{ai} N_{an} N_{bj} N_{bm} p_n \cdot p_m
- \sum \alpha_a N_{an} N_{am} p_n \cdot p_m\,,
\nonumber\\
&\frac{A_1(\alpha,\beta)}{D(\alpha)} =
2 \sum M^{-1}_{ij} \alpha_a \beta_c N_{ai} N_{an} N_{cj} p_n \cdot v
- 2 \sum \beta_c N_{cn} p_n \cdot v\,,
\nonumber\\
&\frac{A_2(\alpha,\beta)}{D(\alpha)} =
\sum M^{-1}_{ij} \beta_c \beta_d N_{ci} N_{dj}\,.
\label{alpha:A12}
\end{align}
The polynomials $D(\alpha)$,
$A(\alpha)$, $A_1(\alpha,\beta)$, $A_2(\alpha,\beta)$
have dimensionality $1/M^{2L}$.
The function $D(\alpha)$ is homogeneous in $\alpha_a$ of degree $L$.
The function $A(\alpha)$ is homogeneous in $\alpha_a$ of degree $L+1$
and linear in $p_n \cdot p_m$.
The function $A_1(\alpha,\beta)$ is linear in $\beta_c$,
of degree $L$ in $\alpha_a$, and linear in $p_n \cdot v$.
The function $A_2(\alpha,\beta)$ is quadratic in $\beta_c$
and of degree $L-1$ in $\alpha_a$;
it does not contain momenta.

It is always possible to calculate (at least) one integration
in~(\ref{alpha:res}).
Let's insert $\delta\left(\sum\alpha_a-\eta\right)d\eta$
under the integral sign,
and make the substitution $\alpha_a = \eta x_a$,
$\beta_c = \eta y_c$.
Then the $\eta$ integral is a $\Gamma$ function:
\begin{equation}
I = \frac{\Gamma\bigl(\sum n_a + \sum n_c - L \frac{d}{2}\bigr)}%
{\prod \Gamma(n_a)\,\prod \Gamma(n_c)}
\int \frac{\prod d x_a\, x_a^{n_a-1}\,\prod d y_c\,y_c^{n_c-1}\,
\delta\bigl(\sum x_a - 1\bigr)}%
{[D(x)]^{d/2} \left[
\frac{A(x)+A_1(x,y)+A_2(x,y)}{D(x)}
+ \sum x_a m_a^2 \right]^{\sum n_a+\sum n_c-Ld/2}}\,.
\label{alpha:Feynman}
\end{equation}
The ordinary Feynman parameters
$x_a$ are dimensionless and vary from 0 to 1;
the HQET Feynman parameters $y_c$ have dimensionality of energy
and vary from 0 to $\infty$.

\subsection{Graph-theoretical rules}
\label{S:graph}

\begin{figure}[ht]
\begin{center}
\begin{picture}(38,23.5)
\put(19,13.5){\makebox(0,0){\includegraphics{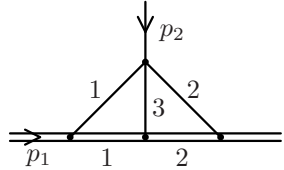}}}
\put(5,3){\makebox(0,0)[t]{$p_1$}}
\put(21,18.5){\makebox(0,0)[l]{$p_2$}}
\put(12.5,11){\makebox(0,0){1}}
\put(25.5,11){\makebox(0,0){2}}
\put(20,8.5){\makebox(0,0)[l]{3}}
\put(14,3){\makebox(0,0)[t]{1}}
\put(24,3){\makebox(0,0)[t]{2}}
\end{picture}
\end{center}
\caption{An HQET vertex diagram}
\label{F:ex}
\end{figure}

The polynomials $D(\alpha)$,
$A(\alpha)$, $A_1(\alpha,\beta)$, $A_2(\alpha,\beta)$
can be extracted directly from the diagram, see~\cite{Z:79}.
We shall formulate the rules and illustrate them
by an example shown in Fig.~\ref{F:ex}.

\textbf{1}. Cut a few light lines so as to get a connected tree,
form the product of $\alpha_a$ of the cut lines.
$D(\alpha)$ is the sum of all such products.
\begin{equation*}
\raisebox{-5mm}{\includegraphics{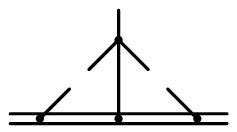}}\qquad
\raisebox{-5mm}{\includegraphics{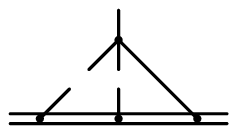}}\qquad
\raisebox{-5mm}{\includegraphics{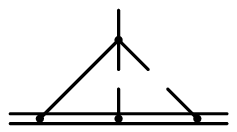}}\qquad
D = \alpha_1 \alpha_2 + \alpha_1 \alpha_3 + \alpha_2 \alpha_3\,.
\end{equation*}

\textbf{2}. Cut a few light lines to get two connected trees
(the heavy line is in one of the two parts).
Form the product of $\alpha_a$ of the cut lines
and multiply it by $(-P^2)$,
where $P$ is the momentum flowing from one connected part
to the other one.
$A(\alpha)$ is the sum of all such terms.
\begin{equation*}
\raisebox{-5mm}{\includegraphics{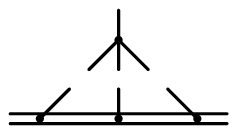}}\qquad
A = - p_2^2 \alpha_1 \alpha_2 \alpha_3\,.
\end{equation*}

\textbf{3}. Cut a single heavy line and a few light ones
to get two connected trees
(now the heavy line enters one connected part
and leaves the other one).
Form the product of $\beta_c$ of the cut heavy line
and $\alpha_a$ of the cut light ones
and multiply by $(-2 P \cdot v)$,
where $P$ is the momentum flowing from the first
connected part to the second one.
$A_1(\alpha,\beta)$ is the sum of all such terms.
\begin{align*}
&\raisebox{-5mm}{\includegraphics{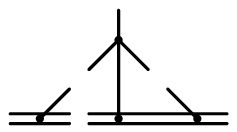}}\qquad
\raisebox{-5mm}{\includegraphics{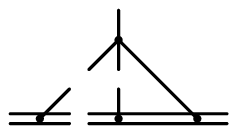}}\qquad
\raisebox{-5mm}{\includegraphics{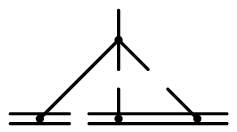}}\\
&\raisebox{-5mm}{\includegraphics{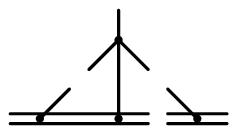}}\qquad
\raisebox{-5mm}{\includegraphics{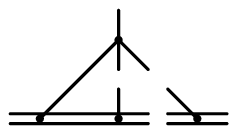}}\qquad
\raisebox{-5mm}{\includegraphics{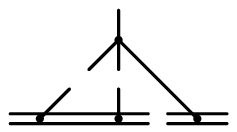}}\\
&A_1 = - 2 p_1 \cdot v\,\beta_1 \alpha_1 (\alpha_2+\alpha_3)
- 2 (p_1+p_2) \cdot v\, \beta_1 \alpha_2 \alpha_3\\
&\hphantom{A_1={}} - 2 (p_1+p_2) \cdot v\, \beta_2 \alpha_2 (\alpha_1+\alpha_3)
- 2 p_1 \cdot v\,\beta_2 \alpha_1 \alpha_3\,.
\end{align*}

\textbf{4}. Cut a few light lines to get
a connected diagram with a single loop
in such a way that this loop contains at least one heavy line.
Sum $\beta_c$ of the heavy lines belonging to the loop,
square the sum, and multiply by $\alpha_a$ of the cut light lines.
Sum all terms.
\begin{align*}
&\raisebox{-5mm}{\includegraphics{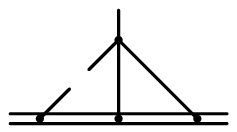}}\qquad
\raisebox{-5mm}{\includegraphics{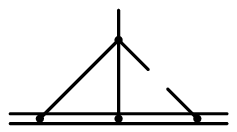}}\qquad
\raisebox{-5mm}{\includegraphics{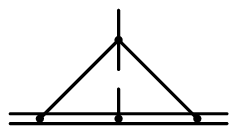}}\\
&A_2 = \alpha_1 \beta_2^2 + \alpha_2 \beta_1^2
+ \alpha_3 (\beta_1+\beta_2)^2\,.
\end{align*}

These rules can be simplified a little.
Suppose $p_1$ is the residual momentum of the incoming heavy line,
and let's route it along the heavy line.
Then the exponent $X$~(\ref{alpha:I})
contains $2p_1\cdot v\,\sum\beta_c$.
We can add this expression to the exponent in~(\ref{alpha:res}),
and then set $p_1=0$ while calculating $A_1(\alpha,\beta)$.

There is an analogy between Feynman diagrams in $\alpha$ representation
and electrical circuits (Table~\ref{T}).
The average momentum flowing through a propagator
corresponds to current.
The first Kirchhoff rule is satisfied:
the sum of momenta flowing into a vertex vanishes.
Light lines are resistors $\alpha_a$,
and heavy lines --- voltage sources $\beta_c v$
(batteries with zero internal resistance,
the voltage does not depend on the current).
The second Kirchhoff rule says that the sum of voltages
along a loop (say, loop $i$) must vanish.
These equations are nothing but the equations
$\sum M_{ij} K_j = Q_i$ which determine the average loop momenta $K_i$
(see~(\ref{alpha:K})).

\begin{table}[ht]
\begin{center}
\begin{tabular}{|l|l|l|}
\hline
& Current & Voltage\\
\hline
\raisebox{-3mm}{\begin{picture}(22,6)
\put(11,4){\makebox(0,0){\includegraphics{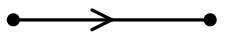}}}
\put(11,0){\makebox(0,0)[b]{$a$}}
\end{picture}}
& $\displaystyle \bar{q}_a = \sum N_{ai} K_i + \sum N_{an} p_n$
& $\displaystyle \alpha_a \bar{q}_a$\\
\hline
\raisebox{-3mm}{\begin{picture}(22,6)
\put(11,4){\makebox(0,0){\includegraphics{grozin_andrey.fig01.eps}}}
\put(11,0){\makebox(0,0)[b]{$c$}}
\end{picture}}
& $\displaystyle \bar{q}_c = \sum N_{ci} K_i + \sum N_{cn} p_n$
& $\displaystyle \beta_c v$\\
\hline
\end{tabular}
\end{center}
\caption{Analogy with electrical circuits}
\label{T}
\end{table}

The Joule heat $\sum \alpha_a \bar{q}_a^2$
plus 2 times the energy consumption by the voltage sources
$\sum \beta_c \bar{q}_c \cdot v$ gives
\begin{equation*}
- \frac{A(\alpha)+A_1(\alpha,\beta)+A_2(\alpha,\beta)}{D(\alpha)}\,.
\end{equation*}
The case $\alpha_a\to0$ or $\beta_c\to0$ corresponds to a short circuit
(line shrinks to a point);
the case $\alpha_a\to\infty$ --- no contact (the line is removed).

Generalization to integrals with numerators
is straightforward~\cite{Z:79}.
Suppose we have a polynomial $\mathcal{P}(q_a,q_c)$
inserted into the numerator of~(\ref{alpha:form}).
Then we can add
\begin{equation*}
2 \sum q_a\cdot\xi_a + 2 \sum q_c\cdot\eta_c
\end{equation*}
to the exponent $X$~(\ref{alpha:I}),
and apply the differential operator
\begin{equation*}
\mathcal{P}\left(\frac{1}{2}\frac{\partial}{\partial\xi_a},
\frac{1}{2}\frac{\partial}{\partial\eta_c}\right)
\end{equation*}
to the result at $\xi_a=0$, $\eta_c=0$.
Before this step, the only difference is the substitution
$\beta_c v\to\beta_c v+\eta_c$,
and the fact that a light line $a$
can be also considered ``heavy'' in $A_1$ and $A_2$ calculations,
with $\xi_a$ playing the role of $\beta v$.
Let's formulate the rules to calculate $A_1$, $A_2$.

\textbf{3$'$}. Cut a single heavy line (say, $c$)
and a few light ones to get two connected trees,
and form the product of $-2(\beta_c v+\eta_c)\cdot P$,
where $P$ is the momentum flowing from the first
connected part to the second one;
multiply this product by $\alpha_a$ of all cut light lines.
Or cut a single light line (say, $a$)
and a few light ones to get two connected trees,
form the product $-2\xi_a\cdot P$,
and multiply it by $\alpha_b$ of these additional cut lines.
Here the first and the second connected parts
are defined by the direction of the momentum
of the first cut line
($q_c$ or $q_a$ for a heavy or light line).
$A_1(\alpha,\beta,\xi,\eta)$ is the sum of all such terms.

\textbf{4$'$}. Cut a few light lines to get
a connected diagram with a single loop.
Sum $\beta_c v+\eta_c$ or $\xi_a$ of the heavy or light lines
belonging to the loop,
square the sum, and multiply by $\alpha_a$ of the cut light lines.
Sum all terms to get $A_2(\alpha,\beta,\xi,\eta)$.

\section{HQET propagator diagrams}
\label{S:Prop}

\subsection{Two loops}
\label{S:L2}

\subsubsection{Diagram 1}
\label{S:L21}

\begin{figure}[t]
\begin{center}
\begin{picture}(84,17)
\put(16,8.5){\makebox(0,0){\includegraphics{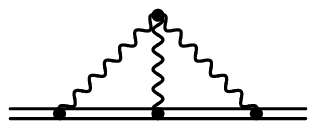}}}
\put(63,8.5){\makebox(0,0){\includegraphics{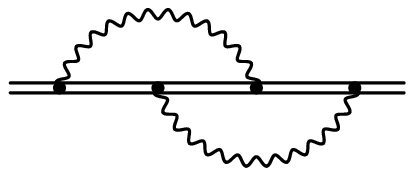}}}
\end{picture}
\end{center}
\caption{Generic topologies of two-loop propagator diagrams}
\label{F:L2t}
\end{figure}

We have calculated the one-loop HQET propagator diagram
by three different methods (Sects.~\ref{S:L1} and~\ref{S:ag}).
Now we shall consider two-loop propagator diagrams in HQET.
There are two generic topologies of such diagrams (Fig.~\ref{F:L2t}).
This means that all other possible topologies can be obtained
from these ones by shrinking some lines.
The method of calculation of these diagrams
has been constructed in~\cite{BG:91}.

\begin{figure}[t]
\begin{center}
\begin{picture}(62,30)
\put(31,16.5){\makebox(0,0){\includegraphics{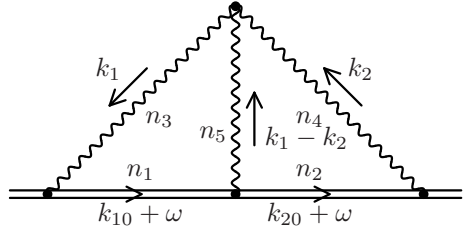}}}
\put(18.5,0){\makebox(0,0)[b]{{$k_{10}+\omega$}}}
\put(41,0){\makebox(0,0)[b]{{$k_{20}+\omega$}}}
\put(14.25,20.75){\makebox(0,0){{$k_1$}}}
\put(47.75,20.75){\makebox(0,0){{$k_2$}}}
\put(35,11.5){\makebox(0,0)[l]{{$k_1-k_2$}}}
\put(18.5,8){\makebox(0,0)[t]{{$n_1$}}}
\put(41,8){\makebox(0,0)[t]{{$n_2$}}}
\put(21,14){\makebox(0,0){{$n_3$}}}
\put(41,14){\makebox(0,0){{$n_4$}}}
\put(30,12){\makebox(0,0)[r]{{$n_5$}}}
\end{picture}
\end{center}
\caption{Diagram 1}
\label{F:L21}
\end{figure}

The first diagram (Fig.~\ref{F:L21}) is
\begin{equation}
- \frac{1}{\pi^d} \int \frac{d^d k_1\,d^d k_2}%
{D_1^{n_1} D_2^{n_2} D_3^{n_3} D_4^{n_4} D_5^{n_5}} =
(-2\omega)^{2d-n_1-n_2-2(n_3+n_4+n_5)} I(n_1,n_2,n_3,n_4,n_5)\,,
\label{L21:def}
\end{equation}
where
\begin{align*}
&D_1=-2(k_{10}+\omega)\,,\quad
D_2=-2(k_{20}+\omega)\,,\\
&D_3=-k_1^2\,,\quad
D_4=-k_2^2\,,\quad
D_5=-(k_1-k_2)^2
\end{align*}
(the power of $-2\omega$ is fixed by dimensionality).
It is symmetric with respect to $(1\leftrightarrow2,3\leftrightarrow4)$,
and vanishes if two adjacent indices are $\le0$.

If $n_5=0$, the diagram is the product of two one-loop ones:
\begin{equation}
I(n_1,n_2,n_3,n_4,0) =
\raisebox{-7.25mm}{\begin{picture}(52,15.5)
\put(26,7.75){\makebox(0,0){\includegraphics{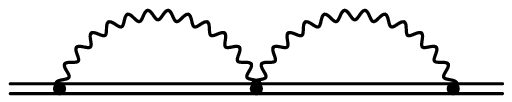}}}
\put(16,0){\makebox(0,0)[b]{{$n_1$}}}
\put(36,0){\makebox(0,0)[b]{{$n_2$}}}
\put(16,15.5){\makebox(0,0)[t]{{$n_3$}}}
\put(36,15.5){\makebox(0,0)[t]{{$n_4$}}}
\end{picture}}
= I(n_1,n_3) I(n_2,n_4)\,.
\label{L21:n5}
\end{equation}
If $n_1=0$, we first calculate the inner massless loop.
The one-loop massless diagram is
\begin{align}
&\frac{1}{i\pi^{d/2}} \int
\frac{d^d k}{\left[-(k+p)^2-i0\right]^{n_1} \left[-k^2-i0\right]^{n_2}}
= (-p^2)^{d/2-n_1-n_2} G(n_1,n_2)\,,
\nonumber\\
&G(n_1,n_2) = \frac{\Gamma(-d/2+n_1+n_2) \Gamma(d/2-n_1) \Gamma(d/2-n_2)}%
{\Gamma(n_1) \Gamma(n_2) \Gamma(d-n_1-n_2)}\,.
\label{L21:G}
\end{align}
This gives the coefficient $G(n_3,n_5)$,
and shifts the power $n_4$ by $n_3+n_5-d/2$:
\begin{align}
&\raisebox{-12mm}{
\begin{picture}(52,25)
\put(26,14){\makebox(0,0){\includegraphics{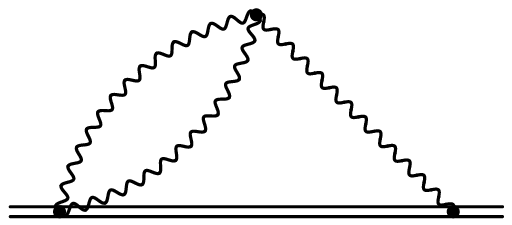}}}
\put(26,0){\makebox(0,0)[b]{$n_2$}}
\put(38.5,15.5){\makebox(0,0){$n_4$}}
\put(10.5,18.5){\makebox(0,0){$n_3$}}
\put(21.5,9){\makebox(0,0){$n_5$}}
\end{picture}} =
\raisebox{-11mm}{
\begin{picture}(32,23)
\put(16,11.5){\makebox(0,0){\includegraphics{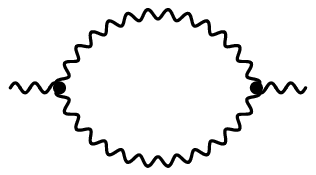}}}
\put(16,0){\makebox(0,0)[b]{$n_5$}}
\put(16,23){\makebox(0,0)[t]{$n_3$}}
\end{picture}} \times
\raisebox{-7.25mm}{
\begin{picture}(32,16)
\put(16,7.75){\makebox(0,0){\includegraphics{grozin_andrey.fig33.eps}}}
\put(16,0){\makebox(0,0)[b]{$n_2$}}
\put(16,16){\makebox(0,0)[t]{$n_4+n_3+n_5-d/2$}}
\end{picture}}\,,
\nonumber\\
&I(0,n_2,n_3,n_4,n_5) = G(n_3,n_5) I(n_2,n_4+n_3+n_5-d/2)
\label{L21:n1}
\end{align}
(the case $n_2=0$ is symmetric).
If $n_3=0$, we first calculate the inner HQET loop.
This gives the coefficient $I(n_1,n_5)$,
and shifts the power $n_2$ by $n_1+2n_5-d$:
\begin{align}
&\raisebox{-11mm}{
\begin{picture}(52,23)
\put(26,11.5){\makebox(0,0){\includegraphics{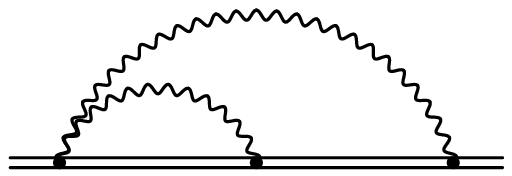}}}
\put(16,0){\makebox(0,0)[b]{$n_1$}}
\put(36,0){\makebox(0,0)[b]{$n_2$}}
\put(26,23){\makebox(0,0)[t]{$n_4$}}
\put(25.5,10.5){\makebox(0,0){$n_5$}}
\end{picture}} =
\raisebox{-7.25mm}{
\begin{picture}(32,16)
\put(16,7.75){\makebox(0,0){\includegraphics{grozin_andrey.fig33.eps}}}
\put(16,0){\makebox(0,0)[b]{$n_1$}}
\put(16,16){\makebox(0,0)[t]{$n_5$}}
\end{picture}} \times
\raisebox{-7.25mm}{
\begin{picture}(32,16)
\put(16,7.75){\makebox(0,0){\includegraphics{grozin_andrey.fig33.eps}}}
\put(16,0){\makebox(0,0)[b]{$n_2+n_1+2n_5-d$}}
\put(16,16){\makebox(0,0)[t]{$n_3$}}
\end{picture}}\,,
\nonumber\\
&I(n_1,n_2,0,n_4,n_5) = I(n_1,n_5) I(n_2+n_1+2n_5-d,n_4)
\label{L21:n3}
\end{align}
(the case $n_4=0$ is symmetric).

But what can we do if all 5 powers of denominators are positive?
We shall use integration by parts~\cite{CT:81}.
Integral of any full derivative over the whole space
of loop momenta is zero.
When applied to the integrand of~(\ref{L21:def}),
the derivative
\begin{equation*}
\frac{\partial}{\partial k_2} \to
\frac{n_2}{D_2}2v
+ \frac{n_4}{D_4}2k_2
+ \frac{n_5}{D_5}2(k_2-k_1)\,.
\end{equation*}
Applying $(\partial/\partial k_2)\cdot k_2$
or $(\partial/\partial k_2)\cdot(k_2-k_1)$
to the integrand, we obtain zero integral.
On the other hand, we can calculate these derivatives explicitly.
Using $2 k_2\cdot v=-D_2-2\omega$, $2(k_2-k_1)\cdot k_2=D_3-D_4-D_5$,
we see that applying these differential operators
is equivalent to inserting
\begin{align*}
&d-n_2-n_5-2n_4 - 2 \omega \frac{n_2}{D_2} + \frac{n_5}{D_5}(D_3-D_4)\,,\\
&d-n_2-n_4-2n_5 + \frac{n_2}{D_2}D_1 + \frac{n_4}{D_4}(D_3-D_5)
\end{align*}
under the integral sign.
These combinations of integrals vanish.
These recurrence relations are usually written as
\begin{align}
\left[ d-n_2-n_5-2n_4 + n_2\2+ + n_5\5+(\3--\4-) \right] I = 0\,,
\label{L21:t1}\\
\left[ d-n_2-n_4-2n_5 + n_2\2+\1- + n_4\4+(\3--\5-) \right] I = 0\,,
\label{L21:t2}
\end{align}
where, for example, $\1-$ lowers $n_1$ by 1 and $\2+$ raises $n_2$ by 1.
Applying $(\partial/\partial k_2)\cdot v$,
we obtain a (less useful) relation
\begin{equation}
\left[ -2n_2\2+ + n_4\4+(\2--1) + n_5\5+(\2--\1-) \right] I = 0\,.
\label{L21:v}
\end{equation}

A useful relation can be obtained from homogeneity
of the integral~(\ref{L21:def}) in $\omega$.
Applying $\omega(d/d\omega)$, we get the same integral times
its dimensionality $2(d-n_3-n_4-n_5)-n_1-n_2$.
On the other hand, we can calculate the derivative explicitly:
\begin{equation}
\left[ 2(d-n_3-n_4-n_5)-n_1-n_2 + n_1\1+ + n_2\2+ \right] I = 0\,.
\label{L21:hom}
\end{equation}
This homogeneity relation is not independent:
it is the sum of the $(\partial/\partial k_2)\cdot k_2$
relation~(\ref{L21:t1}) and its mirror-symmetric
$(\partial/\partial k_1)\cdot k_1$ one.

\begin{figure}[h]
\begin{center}
\begin{picture}(150,50)
\put(25,25){\makebox(0,0){\includegraphics[width=40mm]{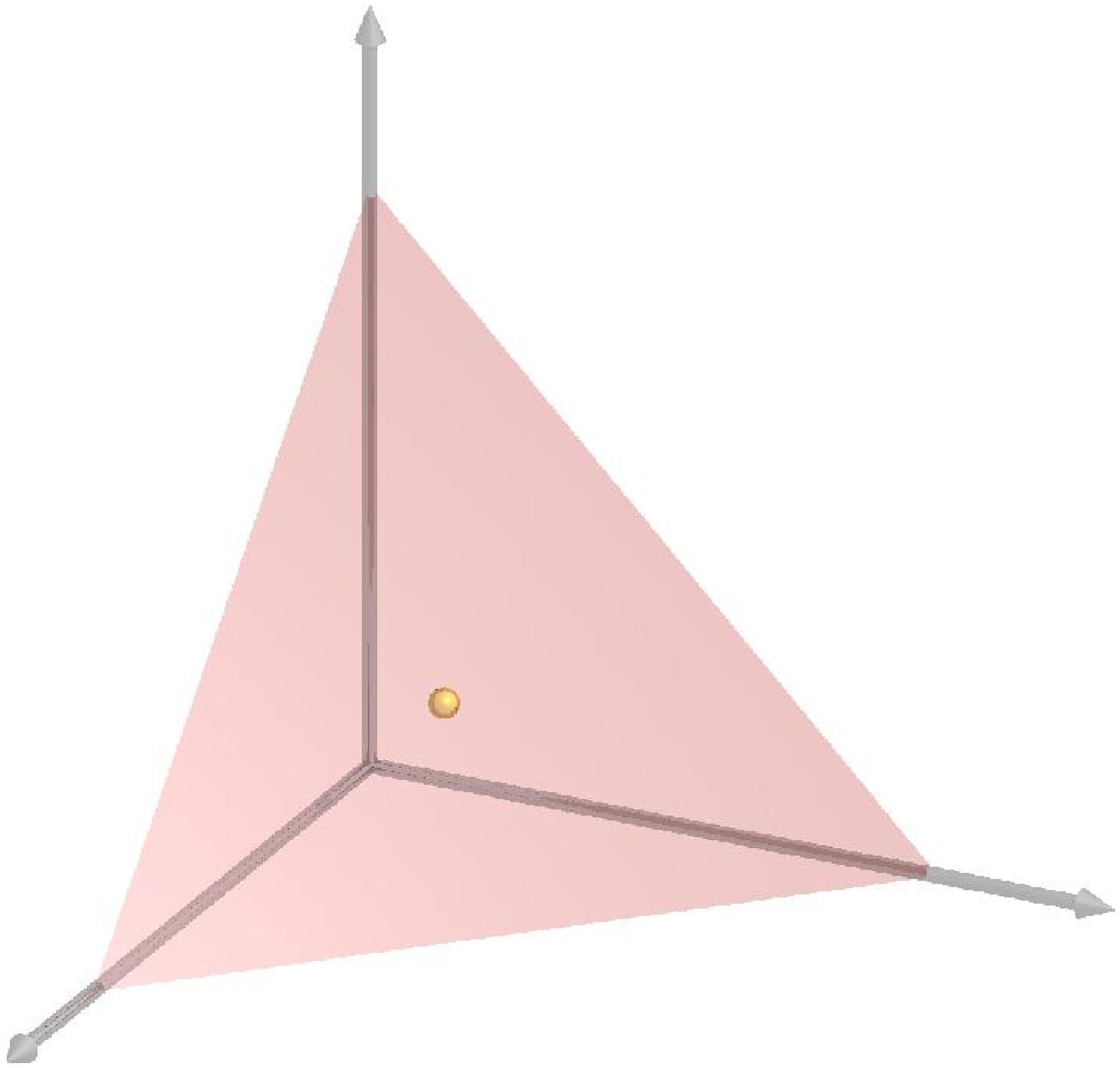}}}
\put(75,25){\makebox(0,0){\includegraphics[width=40mm]{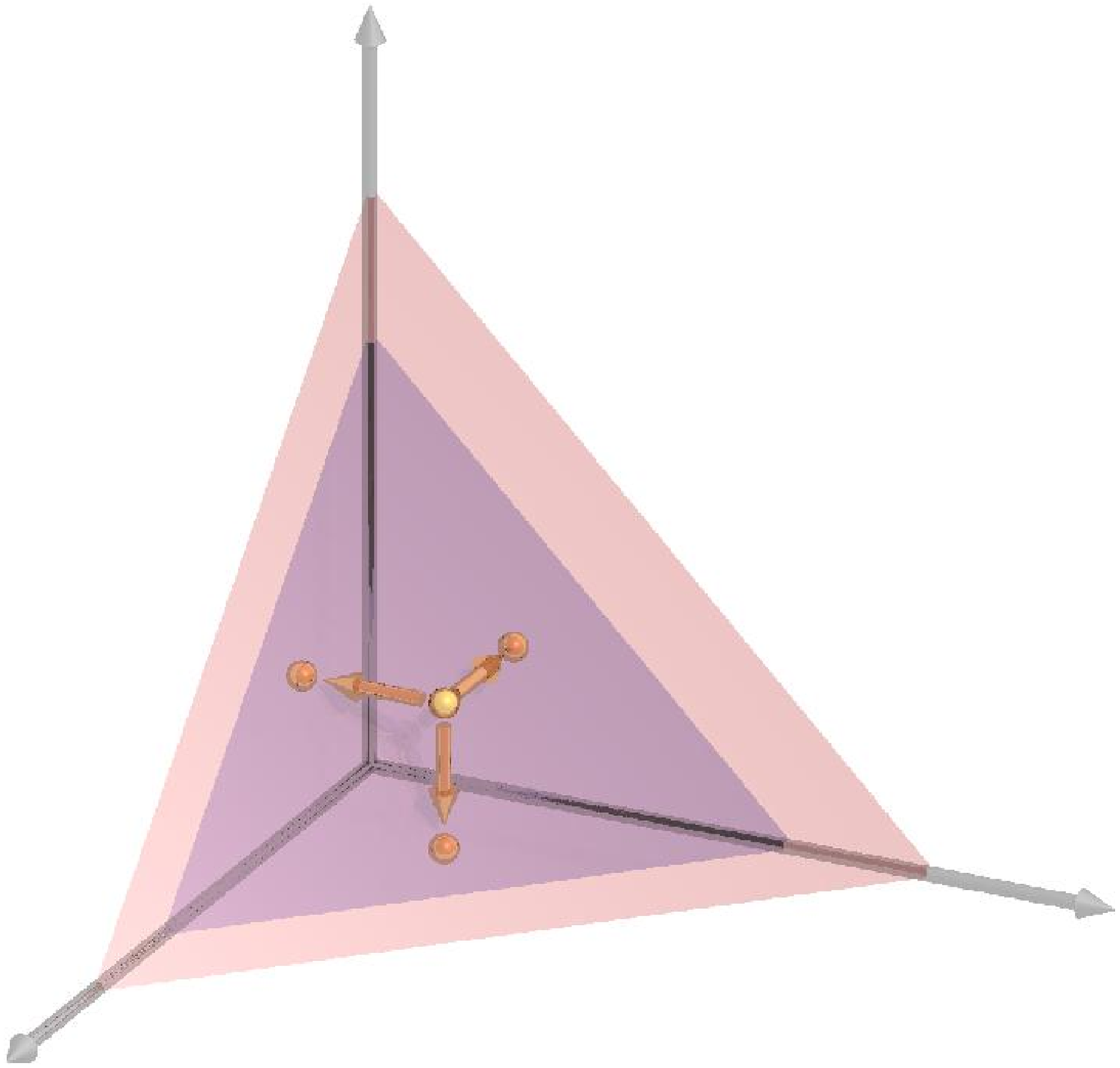}}}
\put(125,25){\makebox(0,0){\includegraphics[width=40mm]{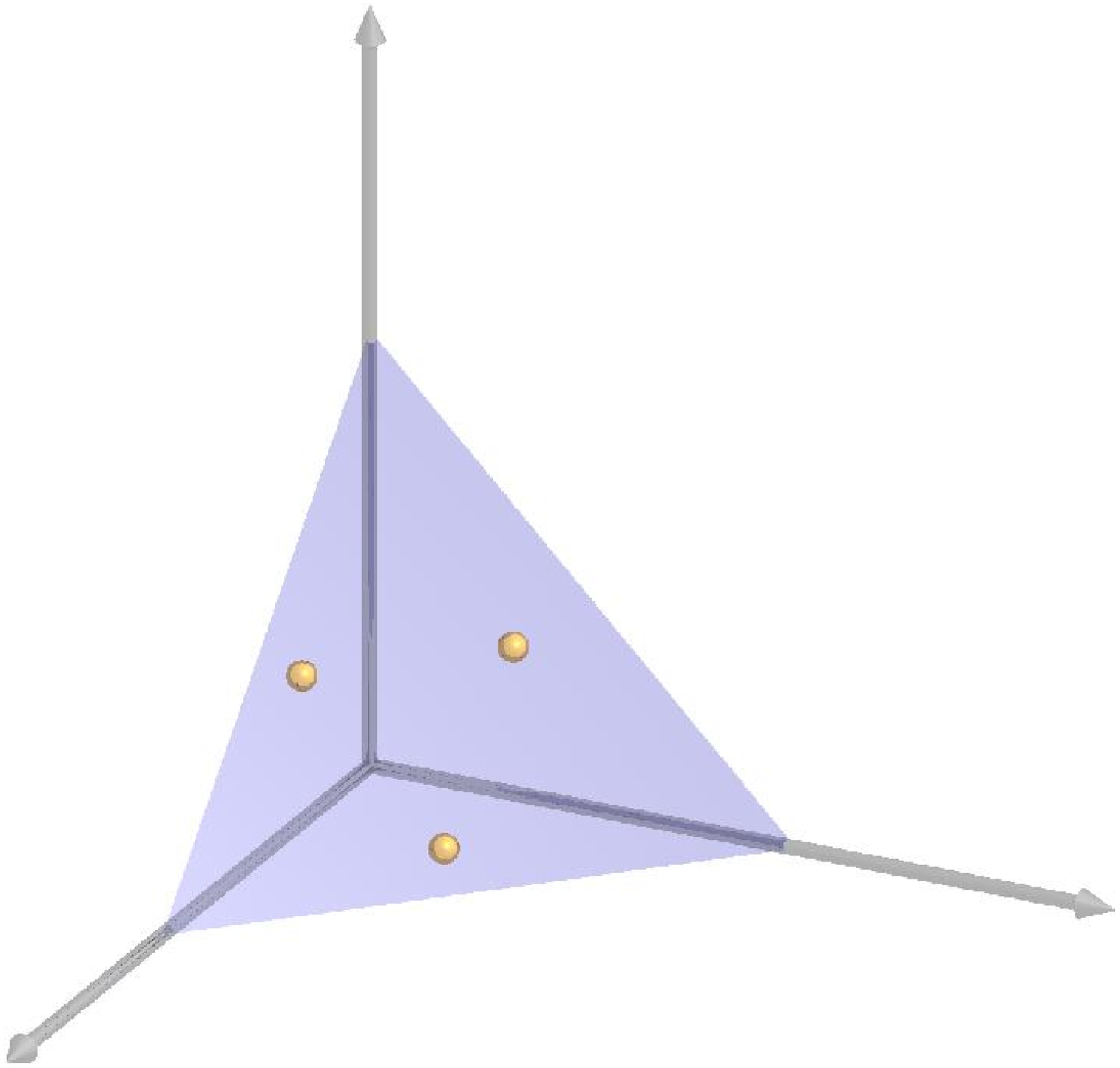}}}
\end{picture}
\end{center}
\caption{A single step of the integration-by-parts reduction}
\label{F:IBP}
\end{figure}

A particularly useful relation can be obtained
by subtracting the $\1-$ shifted homogeneity relation~(\ref{L21:hom})
from the $(\partial/\partial k_2)\cdot(k_2-k_1)$ relation~(\ref{L21:t2}):
\begin{align}
\bigl[& d-n_1-n_2-n_4-2n_5+1
- \bigl(2(d-n_3-n_4-n_5)-n_1-n_2+1\bigr)\1-
\nonumber\\
&{} + n_4\4+(\3--\5-) \bigr] I = 0\,.
\label{L21:David}
\end{align}
Solving it for the $I$ with the unshifted indices,
we obtain an expression for $I(n_1,n_2,n_3,n_4,n_5)$
via 3 integrals:
\begin{equation}
I = \frac{(2(d-n_3-n_4-n_5)-n_1-n_2+1)\1- + n_4\4+(\5--\3-)}%
{d-n_1-n_2-n_4-2n_5+1} I\,.
\label{L21:step}
\end{equation}
Each of them has $n_1+n_3+n_5$ reduced by 1.
Each application of~(\ref{L21:step}) moves us closer to the origin
(Fig.~\ref{F:IBP}).
Therefore, after a finite number of steps,
any integral $I(n_1,n_2,n_3,n_4,n_5)$
will be reduced to the trivial cases
in which one of the indices vanishes.

\subsubsection{Diagram 2}
\label{S:L22}

\begin{figure}[t]
\begin{center}
\begin{picture}(84,50)
\put(42,25){\makebox(0,0){\includegraphics{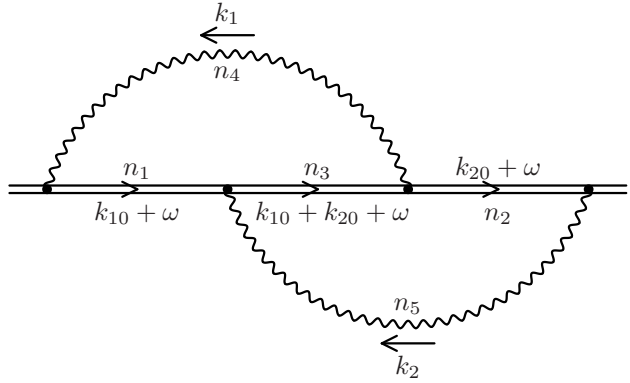}}}
\put(30,50){\makebox(0,0)[t]{{$k_1$}}}
\put(54,0){\makebox(0,0)[b]{{$k_2$}}}
\put(18,23.5){\makebox(0,0)[t]{{$k_{10}+\omega$}}}
\put(44,23.5){\makebox(0,0)[t]{{$k_{10}+k_{20}+\omega$}}}
\put(66,26.5){\makebox(0,0)[b]{{$k_{20}+\omega$}}}
\put(18,26.5){\makebox(0,0)[b]{{$n_1$}}}
\put(42,26.5){\makebox(0,0)[b]{{$n_3$}}}
\put(66,23.5){\makebox(0,0)[t]{{$\vphantom{k}n_2$}}}
\put(30,41.5){\makebox(0,0)[t]{{$n_4$}}}
\put(54,8.5){\makebox(0,0)[b]{{$n_5$}}}
\end{picture}
\end{center}
\caption{Diagram 2}
\label{F:L22}
\end{figure}

The second diagram (Fig.~\ref{F:L22}) is
\begin{equation}
- \frac{1}{\pi^d} \int \frac{d^d k_1\,d^d k_2}%
{D_1^{n_1} D_2^{n_2} D_3^{n_3} D_4^{n_4} D_5^{n_5}} =
(-2\omega)^{2d-n_1-n_2-n_3-2(n_4+n_5)} J(n_1,n_2,n_3,n_4,n_5)\,,
\label{L22:def}
\end{equation}
where
\begin{align*}
&D_1=-2(k_{10}+\omega)\,,\quad
D_2=-2(k_{20}+\omega)\,,\quad
D_3=-2(k_{10}+k_{20}+\omega)\,,\\
&D_4=-k_1^2\,,\quad
D_5=-k_2^2
\end{align*}
(the power of $-2\omega$ is fixed by dimensionality).
It is symmetric with respect to $(1\leftrightarrow2,4\leftrightarrow5)$,
and vanishes if $n_4\le0$ or $n_5\le0$
or two adjacent $n_{1\ldots3}$ are $\le0$.

This integral is trivial if $n_3=0$ or $n_{1,2}=0$.
In general, it has 3 linear denominators and only 2 loop momenta;
therefore, these denominators are linearly dependent:
\begin{equation}
D_1 + D_2 - D_3 = -2\omega\,.
\label{L22:D}
\end{equation}
Inserting this combination under the integral sign, we obtain
\begin{equation}
J = (\1-+\2--\3-) J\,.
\label{L22:parfrac}
\end{equation}
Each application of this recurrence relation reduces $n_1+n_2+n_3$ by 1.
Therefore, after a number of such steps any integral will reduce
to the trivial cases (Fig.~\ref{F:IBP}).

The integral~(\ref{L22:def}) can contain a power of $k_1\cdot k_2$
in the numerator;
this scalar product cannot be expressed via the denominators.
However, this is not a serious problem~\cite{G:00}.

Let's summarize.
All scalar integrals belonging to the two generic topologies
of Fig.~\ref{F:L2t}, with any indices $n_i$
(and with any power of $k_1\cdot k_2$ in the numerator
of the second integral)
can be reduced to linear combinations of two master integrals
\begin{equation}
\raisebox{-3.75mm}{\includegraphics{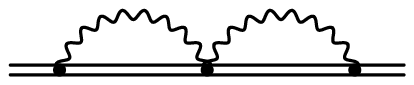}} = I_1^2\,,\quad
\raisebox{-4.5mm}{\includegraphics{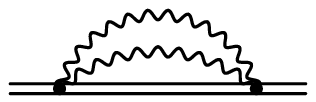}} = I_2\,,
\label{L22:M2}
\end{equation}
with coefficients being rational functions of $d$.
Here the $n$-loop HQET sunset integral is
\begin{equation}
\raisebox{-5.5mm}{\begin{picture}(32,12)
\put(16,6){\makebox(0,0){\includegraphics{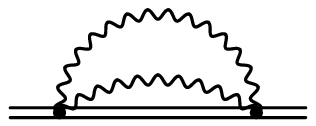}}}
\put(16,7.6666667){\makebox(0,0){{$\cdots$}}}
\end{picture}}
= I_n = \frac{\Gamma(1+2n\varepsilon)\Gamma^n(1-\varepsilon)}%
{(1-n(d-2))_{2n}}\,.
\label{L22:sunset}
\end{equation}
This reduction can be done using integration by parts~\cite{BG:91}
(see also~\cite{G:00}).

\subsection{Three loops}
\label{S:L3}

\subsubsection{Reduction}

\begin{figure}[ht]
\begin{center}
\begin{picture}(112,71.2)
\put(17,64.2){\makebox(0,0){\includegraphics{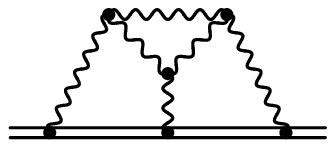}}}
\put(56,64.2){\makebox(0,0){\includegraphics{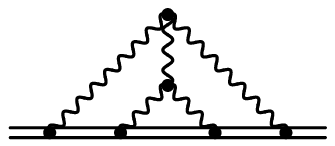}}}
\put(17,47.2){\makebox(0,0){\includegraphics{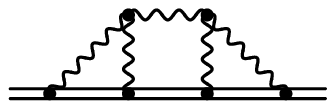}}}
\put(56,47.2){\makebox(0,0){\includegraphics{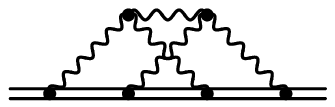}}}
\put(17,29.6){\makebox(0,0){\includegraphics{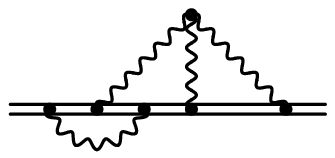}}}
\put(56,29.6){\makebox(0,0){\includegraphics{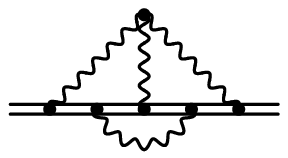}}}
\put(95,27.8){\makebox(0,0){\includegraphics{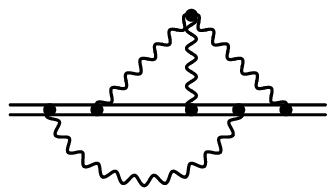}}}
\put(17,7.3){\makebox(0,0){\includegraphics{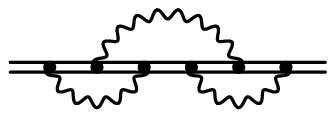}}}
\put(56,7.3){\makebox(0,0){\includegraphics{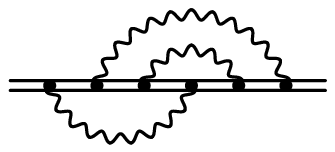}}}
\put(95,9.1){\makebox(0,0){\includegraphics{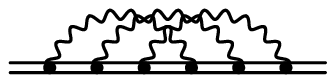}}}
\end{picture}
\end{center}
\caption{Generic topologies of three-loop propagator diagrams}
\label{F:L3t}
\end{figure}

There are 10 generic topologies of three-loop HQET propagator diagrams
(Fig.~\ref{F:L3t}).

All these integrals, with any powers of denominators
and irreducible numerators, can be reduced~\cite{G:00}
to 8 master integrals:
\begin{align}
&\raisebox{-0.25mm}{\includegraphics{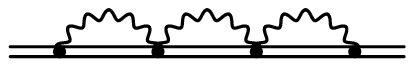}} = I_1^3\,,
\label{L3:b1}\\
&\raisebox{-0.25mm}{\includegraphics{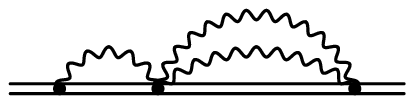}} = I_1 I_2\,,
\label{L3:b2}\\
&\raisebox{-0.25mm}{\includegraphics{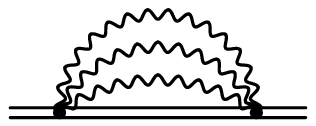}} = I_3\,,
\label{L3:b3}\\
&\raisebox{-0.25mm}{\includegraphics{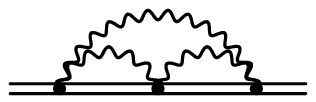}}
= \frac{I_1^2 I(6-2d,1)}{I_2 I(5-2d,1)} I_3
= \frac{3d-7}{2d-5} \frac{I_1^2}{I_2} I_3\,,
\label{L3:b4}\\
&\raisebox{-0.25mm}{\includegraphics{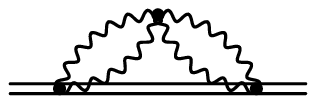}}
= \frac{G_1^2 I(1,4-d)}{G_2 I(1,3-d)} I_3
= - 2 \frac{3d-7}{d-3} \frac{G_1^2}{G_2} I_3\,,
\label{L3:b5}\\
&\raisebox{-0.25mm}{\includegraphics{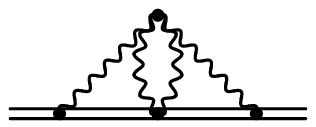}}
= G_1 I(1,1,1,1,2-d/2)\,,
\label{L3:b6}\\
&\raisebox{-8.25mm}{\includegraphics{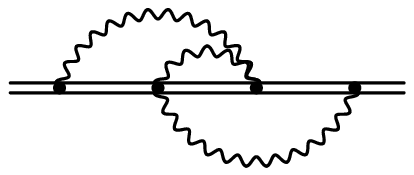}}
= I_1 J(1,1,3-d,1,1)\,,
\label{L3:b7}\displaybreak\\
&\raisebox{-0.25mm}{\includegraphics{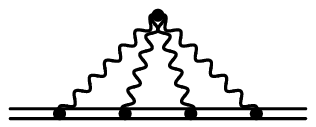}}
= B_8\,,
\label{L3:b8}
\end{align}
using integration by parts.
Here the $n$-loop HQET sunset $I_n$ is defined by~(\ref{L22:sunset}),
and the $n$-loop massless sunset is
\begin{equation}
\raisebox{-8mm}{\begin{picture}(32,17)
\put(16,8.5){\makebox(0,0){\includegraphics{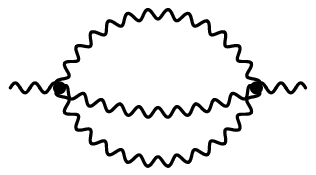}}}
\put(16,11){\makebox(0,0){$\cdots$}}
\end{picture}} =
G_n = \frac{1}{\left(n+1-n\frac{d}{2}\right)_n
\left((n+1)\frac{d}{2}-2n-1\right)_n}
\frac{\Gamma(1+n\varepsilon)\Gamma^{n+1}(1-\varepsilon)}%
{\Gamma(1-(n+1)\varepsilon)}\,.
\label{L3:sunset}
\end{equation}
This reduction algorithm has been implemented
as a \texttt{REDUCE} package \texttt{Grinder}~\cite{G:00}.
It is analogous to the massless package Mincer~\cite{Mincer}.
The first 5 master integrals can be easily expressed
via $\Gamma$ functions, exactly in $d$ dimensions.
The next two ones reduce to two-loop ones with a single
$\varepsilon$-dependent index%
\footnote{\texttt{Grinder} uses $B_4=I_3 I_1^2/I2$ and $B_5=I_3 G_1^2/G_2$
as elements of its basis instead of~(\ref{L3:b4}) and~(\ref{L3:b5}).}
(Sects.~\ref{S:J} and~\ref{S:I}).
The last one is truly three-loop (Sect.~\ref{S:Inv}).

\subsubsection{$J(1,1,n,1,1)$}
\label{S:J}

\begin{figure}[h]
\begin{center}
\begin{picture}(118,28)
\put(27,16.5){\makebox(0,0){\includegraphics{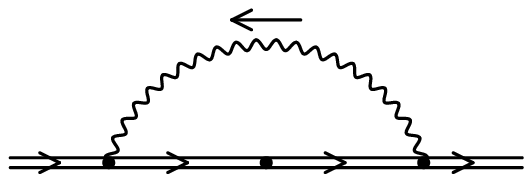}}}
\put(19,4){\makebox(0,0)[b]{$k_0+\omega_1$}}
\put(36,4){\makebox(0,0)[b]{$k_0+\omega_2$}}
\put(27,28){\makebox(0,0)[t]{$k$}}
\put(6,4){\makebox(0,0)[b]{$\omega_1$}}
\put(48,4){\makebox(0,0)[b]{$\omega_2$}}
\put(19,13){\makebox(0,0)[t]{$n_1$}}
\put(36,13){\makebox(0,0)[t]{$n_2$}}
\put(27,17){\makebox(0,0)[b]{$n_3$}}
\put(27,0){\makebox(0,0)[b]{a}}
\put(91,15){\makebox(0,0){\includegraphics{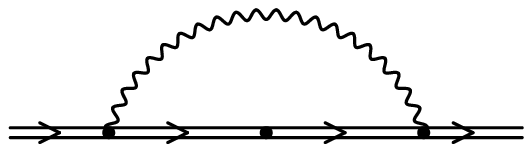}}}
\put(91,4){\makebox(0,0)[b]{$0$}}
\put(75,4){\makebox(0,0)[b]{$-v t_1$}}
\put(107,4){\makebox(0,0)[b]{$v t_2$}}
\put(83,13){\makebox(0,0)[t]{$n_1$}}
\put(99,13){\makebox(0,0)[t]{$n_2$}}
\put(91,17){\makebox(0,0)[b]{$n_3$}}
\put(91,0){\makebox(0,0)[b]{b}}
\end{picture}
\end{center}
\caption{One-loop diagram}
\label{F:BBG}
\end{figure}

Here we shall calculate the integral $J$~(\ref{L22:def})
(Fig.~\ref{F:L22}) for arbitrary powers of denominators.
To this end, we shall first consider the one-loop diagram
with two different residual energies $\omega_1$ and $\omega_2$
(Fig.~\ref{F:BBG}a):
\begin{equation}
I = \frac{1}{i\pi^{d/2}} \int \frac{d^d k}{D_1^{n_1} D_2^{n_2} D_3^{n_3}}\,,
\label{J:Idef}
\end{equation}
where
\begin{equation*}
D_1 = -2(k_0+\omega_1)\,,\quad
D_2 = -2(k_0+\omega_2)\,,\quad
D_3 = -k^2\,.
\end{equation*}
If $n_{1,2}$ are integer, this integral can be easily calculated
by partial fraction decomposition (Sect.~\ref{S:Ren}).

Closely following Sect.~\ref{S:L1}, we first integrate in $d^{d-1}\vec{k}$:
\begin{align*}
I &= \frac{\Gamma\bigl(n_3-\frac{d-1}{2}\bigr)}{\pi^{1/2} \Gamma(n_3)}
\int \frac{\bigl(k_{E0}^2\bigr)^{(d-1)/2-n_3} d k_{E0}}%
{\bigl(- 2 \omega_1 - 2 i k_{E0}\bigr)^{n_1}
\bigl(- 2 \omega_2 - 2 i k_{E0}\bigr)^{n_2}}\\
&= 2
\frac{\Gamma\bigl(n_3-\frac{d-1}{2}\bigr)}{\pi^{1/2} \Gamma(n_3)}
\cos\left[\pi\left(\frac{d}{2}-n_3\right)\right]
\int_0^\infty \frac{k^{d-1-2n_3} d k}%
{(2k-2\omega_1)^{n_1}(2k-2\omega_2)^{n_2}}\,.
\end{align*}
We obtain~\cite{BBG:93}
\begin{equation}
I = I(n_1+n_2,n_3)
{}_2F_1\left(\left.
\begin{array}{c}
n_1,n_1+n_2+2n_3-d\\
n_1+n_2
\end{array}
\right| 1 - \frac{\omega_1}{\omega_2} \right)
(-2\omega_2)^{d-n_1-n_2-2n_3}\,.
\label{J:I}
\end{equation}
One can easily check that this result is symmetric with respect to
$(\omega_1\leftrightarrow\omega_2,n_1\leftrightarrow n_2)$,
using properties of hypergeometric function.
If $\omega_1=\omega_2$, it reduces to $I(n_1+n_2,n_3)$.

Let's also calculate this integral using $\alpha$ representation
(Sect.~\ref{S:ag}):
\begin{align*}
I = \frac{1}{\Gamma(n_1) \Gamma(n_2) \Gamma(n_3)}
\int &d\alpha\,\alpha^{n_3-1}\,
d\beta_1\,\beta_1^{n_1-1}\,d\beta_2\,\beta_2^{n_2-1}\,\alpha^{-d/2}\\
&\times\exp\left[-\frac{(\beta_1+\beta_2)^2}{\alpha}
+2(\omega_1\beta_1+\omega_2\beta_2)\right]\,.
\end{align*}
Now we make the substitution $\beta_{1,2}=\alpha y_{1,2}$
and integrate in $\alpha$:
\begin{equation*}
I = \frac{\Gamma\bigl(n_1+n_2+n_3-\frac{d}{2}\bigr)}%
{\Gamma(n_1) \Gamma(n_2) \Gamma(n_3)}
\int d y_1\,y_1^{n_1-1}\,d y_2\,y_2^{n_2-1}
\bigl[(y_1+y_2)^2-2(\omega_1 y_1+\omega_2 y_2)\bigr]^{d/2-n_1-n_2-n_3}\,.
\end{equation*}
After the substitution $y_1=yx$, $y_2=y(1-x)$,
the integral in $y$ can be taken:
\begin{align}
I = &\frac{\Gamma\bigl(\frac{d}{2}-n_3\bigr)
\Gamma\bigl(n_1+n_2+2n_3-\frac{d}{2}\bigr)}%
{\Gamma(n_1) \Gamma(n_2) \Gamma(n_3)}
\nonumber\\
&\times\int_0^1 dx\,x^{n_1-1}\,(1-x)^{n_2-1}\,
\left[ - 2\omega_1 x - 2\omega_2 (1-x) \right]^{d-n_1-n_2-2n_3}\,.
\label{J:Ix}
\end{align}
And we again obtain~(\ref{J:I}).

Finally, we shall derive the same result in coordinate space
(Fig.~\ref{F:BBG}b, see Sect.~\ref{S:L1}):
\begin{equation*}
I = - \frac{1}{4} \frac{\Gamma\bigl(\frac{d}{2}-n_3\bigr)}%
{\Gamma(n_1) \Gamma(n_2) \Gamma(n_3)}
\int d t_1\,d t_2\,e^{i(\omega_1 t_1 + \omega_2 t_2)}
\left(\frac{i t_1}{2}\right)^{n_1-1}
\left(\frac{i t_2}{2}\right)^{n_2-1}
\left(\frac{i(t_1+t_2)}{2}\right)^{2n_3-d1}\,.
\end{equation*}
The substitution $t_1=tx$, $t_2=t(1-x)$ reduces this expression
to~(\ref{J:Ix}).

Now we return to our main problem ---
calculating $J=J(n_1,n_2,n_3,n_4,n_5)$
(Fig.~\ref{F:L22}) with arbitrary indices.
We set $-2\omega=1$;
the power of $-2\omega$ can be reconstructed by dimensionality.
Substituting the one-loop subdiagram~(\ref{J:I}), we have
\begin{align*}
J =& \frac{I(n_1+n_3,n_4)}{i\pi^{d/2}}
\int \frac{d^d k}{(-k^2)^{n_5} (1-2k_0)^{n_2}}\\
&{}\times (1-2k_0)^{d-n_1-n_3-2n_4}
{}_2 F_1 \left(\left.
\begin{array}{c}
n_1,n_1+n_3+2n_4-d\\
n_1+n_3
\end{array}
\right| \frac{-2k_0}{1-2k_0} \right)\\
=& \frac{I(n_1+n_3,n_4) \Gamma\bigl(n_5-\frac{d-1}{2}\bigr)}%
{\pi^{d/2} \Gamma(n_5)}
\int_{-\infty}^{+\infty}
\frac{(k_{E0}^2)^{(d-1)/2-n_5} d k_{E0}}{(1-2ik_{E0})^{n_1+n_2+n_3+2n_4-d}}\\
&{}\times{}_2 F_1 \left(\left.
\begin{array}{c}
n_1,n_1+n_3+2n_4-d\\
n_1+n_3
\end{array}
\right| \frac{-2ik_{E0}}{1-2ik_{E0}} \right)\,.
\end{align*}
We can deform the integration contour
(Fig.~\ref{F:Contour0}, $k_{E0}=iz/2$):
\begin{align*}
J =& \frac{I(n_1+n_3,n_4) \Gamma\bigl(n_5-\frac{d-1}{2}\bigr)}%
{2^{d-2n_5-1} \pi^{d/2} \Gamma(n_5)}
\cos\left[\pi\left(\frac{d}{2}-n_5\right)\right]\\
&{}\times
\int_0^\infty \frac{z^{d-2n_5-1} dz}{(z+1)^{n_1+n_2+n_3+2n_4-d}}
{}_2 F_1 \left(\left.
\begin{array}{c}
n_1,n_1+n_3+2n_4-d\\
n_1+n_3
\end{array}
\right| \frac{z}{z+1} \right)\,.
\end{align*}
Now we substitute the series
\begin{equation*}
{}_2 F_1 \left( \left.
\begin{array}{c}a,b\\c\end{array}
\right| x \right) =
\frac{\Gamma(c) \Gamma(b)}{\Gamma(a)}
\sum_{n=0}^\infty
\frac{\Gamma(n+a) \Gamma(n+b)}{\Gamma(n+1) \Gamma(n+c)} x^n\,,
\end{equation*}
and integrate term by term.
The result is
\begin{align*}
J =& \frac{I(n_1+n_3,n_4) \Gamma\bigl(n_5-\frac{d-1}{2}\bigr)
\Gamma(d-2n_5) \Gamma(n_1+n_2+n_3+2n_4+2n_5-2d)}%
{2^{d-2n_5-1} \pi^{d/2} \Gamma(n_5) \Gamma(n_1+n_2+n_3+2n_4-d)}
\cos\left[\pi\left(\frac{d}{2}-n_5\right)\right]\\
&{}\times {}_3 F_2 \left( \left.
\begin{array}{c}
n_1,n_1+n_3+2n_4-d,d-2n_5\\
n_1+n_3,n_1+n_2+n_3+2n_4-d
\end{array}
\right| 1 \right)\,.
\end{align*}
Using~(\ref{L1:Gamma}), we can simplify this result:
\begin{align}
&J(n_1,n_2,n_3,n_4,n_5) =
\nonumber\\
&\frac{\Gamma\bigl(\frac{d}{2}-n_4\bigr) \Gamma\bigl(\frac{d}{2}-n_5\bigr)
\Gamma(n_1+n_3+2n_4-d) \Gamma(n_1+n_2+n_3+2n_4+2n_5-2d)}%
{\Gamma(n_4) \Gamma(n_5) \Gamma(n_1+n_3)
\Gamma(n_1+n_2+n_3+2n_4-d)}
\nonumber\\
&{}\times {}_3 F_2 \left( \left.
\begin{array}{c}
n_1,n_1+n_3+2n_4-d,d-2n_5\\
n_1+n_3,n_1+n_2+n_3+2n_4-d
\end{array}
\right| 1 \right)\,.
\label{J:J}
\end{align}
It was first derived in coordinate space~\cite{G:00,G:07}.
Checking the symmetry $(n_1\leftrightarrow n_2,n_4\leftrightarrow n_5)$
requires using some ${}_3 F_2$ identities.

\subsubsection{$I(1,1,1,1,n)$}
\label{S:I}

This diagram has been calculated in~\cite{BB:94}
using Gegenbauer polynomial technique in coordinate space~\cite{CKT:80}:
\begin{align}
&\raisebox{-5.5mm}{\begin{picture}(32,12)
\put(16,6){\makebox(0,0){\includegraphics{grozin_andrey.fig27.eps}}}
\put(18,5){\makebox(0,0){$n$}}
\end{picture}} = I(1,1,1,1,n) =
\frac{\Gamma\left(\frac{d}{2}-1\right)\Gamma\left(\frac{d}{2}-n-1\right)}%
{\Gamma(d-2)}
\nonumber\\
&{}\times\Biggl[ 2
\frac{\Gamma(2n-d+3)\Gamma(2n-2d+6)}{(n-d+3)\Gamma(3n-2d+6)}
{}_3 F_2 \left( \left.
\begin{array}{c}n-d+3,n-d+3,2n-2d+6\\n-d+4,3n-2d+6\end{array}
\right|1\right)
\nonumber\\
&\qquad{} - \Gamma(d-n-2) \Gamma^2(n-d+3) \Biggr]\,.
\label{I:In}
\end{align}
Some details of this method are discussed in~\cite{G:07}.

\subsubsection{Inversion}
\label{S:Inv}

The last three-loop master integral has been calculated~\cite{CM:02}
using inversion.
We shall first consider inversion relations at one and two loops.
The one-loop massive on-shell integral defined by
\begin{equation}
\int \frac{d^d k}%
{\left[m^2-(k+mv)^2-i0\right]^{n_1} \left[-k^2-i0\right]^{n_2}}
= i \pi^{d/2} m^{d-2(n_1+n_2)} M(n_1,n_2)
\label{Inv:M}
\end{equation}
can be written in terms of the dimensionless Euclidean momentum $K=k_E/m$:
\begin{equation*}
\int \frac{d^d K}{(K^2-2iK_0)^{n_1}(K^2)^{n_2}} =
\pi^{d/2} M(n_1,n_2)\,.
\end{equation*}
Similarly, the one-loop HQET propagator integral~(\ref{L1:def})
expressed via $K=k_E/(-2\omega)$ is
\begin{equation*}
\int \frac{d^d K}{(1-2iK_0)^{n_1}(K^2)^{n_2}} =
\pi^{d/2} I(n_1,n_2)\,.
\end{equation*}
Inversion $K=K'/K^{\prime2}$ transforms the massive on-shell denominator
into the HQET one:
\begin{equation*}
K^2 - 2 i K_0 = \frac{1 - 2 i K_0'}{K^{\prime2}}\,.
\end{equation*}
Therefore,
\begin{align}
&\raisebox{-7.25mm}{
\begin{picture}(32,16)
\put(16,7.75){\makebox(0,0){\includegraphics{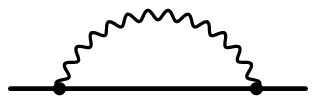}}}
\put(16,0){\makebox(0,0)[b]{{$n_1$}}}
\put(16,16){\makebox(0,0)[t]{{$\vphantom{d}n_2$}}}
\end{picture}} =
\raisebox{-7.25mm}{
\begin{picture}(32,16)
\put(16,7.75){\makebox(0,0){\includegraphics{grozin_andrey.fig33.eps}}}
\put(16,0){\makebox(0,0)[b]{$n_1$}}
\put(16,16){\makebox(0,0)[t]{$d-n_1-n_2$}}
\end{picture}}\,,
\nonumber\\
&M(n_1,n_2) = I(n_1,d-n_1-n_2) =
\frac{\Gamma(d-n_1-2n_2)\Gamma(-d/2+n_1+n_2)}%
{\Gamma(n_1)\Gamma(d-n_1-n_2)}\,.
\label{Inv:L1}
\end{align}

Similarly, at two loops we obtain~\cite{BG:95a}
\begin{equation}
\raisebox{-11.25mm}{
\begin{picture}(42,23)
\put(21,11.5){\makebox(0,0){\includegraphics{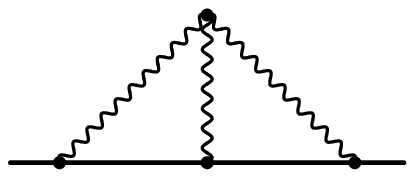}}}
\put(13.5,0){\makebox(0,0)[b]{{$n_1$}}}
\put(28.5,0){\makebox(0,0)[b]{{$n_2$}}}
\put(13,13){\makebox(0,0)[r]{{$n_3$}}}
\put(29,13){\makebox(0,0)[l]{{$n_4$}}}
\put(22,10){\makebox(0,0)[l]{{$n_5$}}}
\end{picture}} =
\raisebox{-11.25mm}{
\begin{picture}(52,23)
\put(26,11.5){\makebox(0,0){\includegraphics{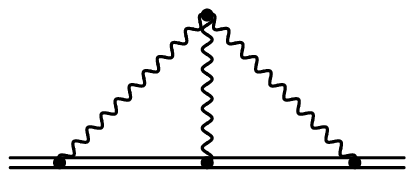}}}
\put(18.5,0){\makebox(0,0)[b]{{$n_1$}}}
\put(33.5,0){\makebox(0,0)[b]{{$n_2$}}}
\put(20,15){\makebox(0,0)[r]{{$d-n_1-n_3$}}}
\put(16,11){\makebox(0,0)[r]{{${}-n_5$}}}
\put(32,15){\makebox(0,0)[l]{{$d-n_2-n_4$}}}
\put(36,11){\makebox(0,0)[l]{{${}-n_5$}}}
\put(27,10){\makebox(0,0)[l]{{$n_5$}}}
\end{picture}}\,.
\label{Inv:L2}
\end{equation}
This relation is less useful, because the HQET diagram
in the right-hand side contains two non-integer indices.

At three loops we have~\cite{G:07}
\begin{align}
&\raisebox{-8.5mm}{\begin{picture}(42,23)
\put(21,11.5){\makebox(0,0){\includegraphics{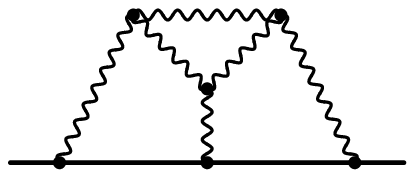}}}
\put(13.5,0){\makebox(0,0)[b]{$n_1$}}
\put(28.5,0){\makebox(0,0)[b]{$n_2$}}
\put(8.5,11.5){\makebox(0,0)[r]{$n_3$}}
\put(33,11.5){\makebox(0,0)[l]{$n_4$}}
\put(22,7.75){\makebox(0,0)[l]{$n_5$}}
\put(21,20){\makebox(0,0)[b]{$n_6$}}
\put(17.7,13){\makebox(0,0)[r]{$n_7$}}
\put(25,13){\makebox(0,0)[l]{$n_8$}}
\end{picture}} =
\raisebox{-8.5mm}{\begin{picture}(58,23)
\put(29,11.5){\makebox(0,0){\includegraphics{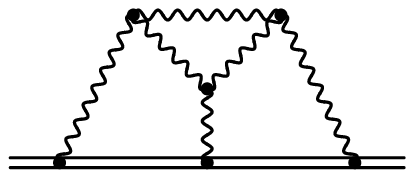}}}
\put(21.5,0){\makebox(0,0)[b]{$n_1$}}
\put(36.5,0){\makebox(0,0)[b]{$n_2$}}
\put(19,17){\makebox(0,0)[r]{$d-n_1-n_3$}}
\put(17,13){\makebox(0,0)[r]{${}-n_5-n_7$}}
\put(38.5,17){\makebox(0,0)[l]{$d-n_2-n_4$}}
\put(40,13){\makebox(0,0)[l]{${}-n_5-n_8$}}
\put(30,7.75){\makebox(0,0)[l]{$n_5$}}
\put(29,20){\makebox(0,0)[b]{$d-n_6-n_7-n_8$}}
\put(25.7,13){\makebox(0,0)[r]{$n_7$}}
\put(33,13){\makebox(0,0)[l]{$n_8$}}
\end{picture}}
\nonumber\\
&\raisebox{-8.5mm}{\begin{picture}(42,18)
\put(21,9){\makebox(0,0){\includegraphics{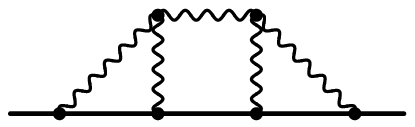}}}
\put(11,0){\makebox(0,0)[b]{$n_1$}}
\put(21,0){\makebox(0,0)[b]{$n_3$}}
\put(31,0){\makebox(0,0)[b]{$n_2$}}
\put(9.3,9){\makebox(0,0)[r]{$n_4$}}
\put(32.2,9){\makebox(0,0)[l]{$n_5$}}
\put(16.8,9){\makebox(0,0)[l]{$n_6$}}
\put(25.2,9){\makebox(0,0)[r]{$n_7$}}
\put(21,15){\makebox(0,0)[b]{$n_8$}}
\end{picture}} =
\raisebox{-8.5mm}{\begin{picture}(58,18)
\put(29,9){\makebox(0,0){\includegraphics{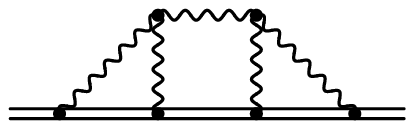}}}
\put(19,0){\makebox(0,0)[b]{$n_1$}}
\put(29,0){\makebox(0,0)[b]{$n_3$}}
\put(39,0){\makebox(0,0)[b]{$n_2$}}
\put(19.5,13){\makebox(0,0)[r]{$d-n_1-n_4$}}
\put(15.5,8.5){\makebox(0,0)[r]{${}-n_6$}}
\put(38,13){\makebox(0,0)[l]{$d-n_2-n_5$}}
\put(41,8.5){\makebox(0,0)[l]{${}-n_7$}}
\put(24.8,9){\makebox(0,0)[l]{$n_6$}}
\put(33.2,9){\makebox(0,0)[r]{$n_7$}}
\put(29,15){\makebox(0,0)[b]{$d-n_3-n_6-n_7-n_8$}}
\end{picture}}
\label{H3:inv}
\end{align}
In particular, the HQET ladder diagram
with all indices $n_i=1$ is convergent;
its value at $d=4$ is related~\cite{CM:02}
to a massive on-shell diagram
\begin{equation}
\raisebox{-4.5mm}{\includegraphics{grozin_andrey.fig44.eps}} =
\raisebox{-4.5mm}{\includegraphics{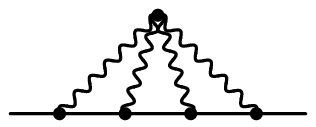}} =
- 5 \zeta_5 + 12 \zeta_2 \zeta_3\,.
\label{H3:inv4}
\end{equation}
by the second inversion relation.
This is one of the on-shell three-loop master integrals,
and its value at $d=4$ is known~\cite{LR:96,MR:00}.
Calculating this ladder diagram with \texttt{Grinder}:
\begin{align}
&\raisebox{-4.5mm}{\includegraphics{grozin_andrey.fig44.eps}} =
4 \frac{(d-3)^2}{(d-4)^2} I_1^3
- \frac{136}{3} \frac{(d-3)(2d-5)(2d-7)}{(d-4)^3} I_1 I_2
\nonumber\\
&{} + 2 \frac{(3d-7)(3d-8)(81d^3-891d^2+3266d-3988)}{(d-4)^4 (2d-7)} I_3
\nonumber\\
&{} + 9 \frac{(d-3)(3d-7)(3d-8)(3d-10)(3d-11)}{(d-4)^3 (2d-5)(2d-7)}
\frac{I_1^2}{I_2} I_3
+ 8 \frac{(d-3)(3d-7)(3d-11)}{(d-4)^3} \frac{G_1^2}{G_2} I_3
\nonumber\\
&{} - \frac{3}{2} \frac{(d-3)(3d-10)}{(d-4)^2}
G_1 I\bigl(1,1,1,1,2-\tfrac{d}{2}\bigr)
- \frac{3d-11}{d-4} B_8\,,
\label{H3:Grinder}
\end{align}
and solving for the most difficult HQET three-loop master integral 
$B_8$~(\ref{L3:b8}),
we obtain the $\varepsilon$ expansion of this integral
up to $\mathcal{O}(\varepsilon)$.
This concludes the investigation of three-loop master integrals,
and allows one to solve three-loop propagator problems in HQET
up to terms $\mathcal{O}(1)$.

\subsubsection{Applications}
\label{S:app1}

Using this technique, the HQET heavy-quark propagator
has been calculated up to three loops~\cite{CG:03},
and the heavy-quark field anomalous dimension
(obtained earlier by a completely different method~\cite{MR:00})
has been confirmed.
The anomalous dimension of the HQET heavy--light quark current
has been calculated~\cite{CG:03}.
The correlator of two heavy--light currents has been found,
up to three loops, including light-quark mass corrections
of order $m$ and $m^2$~\cite{CG2}.
The quark-condensate contribution to this correlator
has been also calculated up to three loops~\cite{CG2}.
Its ultraviolet divergence yields the difference
of twice the anomalous dimension of the heavy-quark current
and the that of the quark condensate,
thus providing a completely independent confirmation
of the result obtained in~\cite{CG:03}.
The gluon-condensate contribution has been calculated
up to two loops~\cite{CG2}
(at one loop it vanishes).

\section{On-shell HQET propagator diagrams with mass}
\label{S:M}

\subsection{Two loops}
\label{S:M2}

On-shell HQET propagator diagrams vanish if all flavours
(except the HQET one) are considered massless,
because loop integrals contain no scale.
If there is a massive flavour ($c$ in the $b$-quark HQET),
such diagrams are non-zero.
They first appear at two loops.
They are used, e.g., to calculate on-shell renormalization constants
in HQET.

\begin{figure}[t]
\begin{center}
\begin{picture}(52,26)
\put(26,14.5){\makebox(0,0){\includegraphics{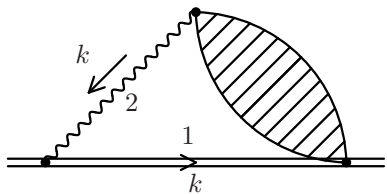}}}
\put(26,3){\makebox(0,0)[t]{$k$}}
\put(11,19){\makebox(0,0){$k$}}
\put(25,7){\makebox(0,0)[b]{1}}
\put(17.5,12.5){\makebox(0,0){2}}
\end{picture}
\end{center}
\caption{Diagram with a single HQET line}
\label{F:blob}
\end{figure}

Let's first consider~\cite{GSS:06}
a class of such integrals (Fig.~\ref{F:blob})
\begin{equation}
F(n_1,n_2) = \int \frac{f(k^2)\,d^d k}{D_1^{n_1} D_2^{n_2}}\,,
\quad\text{where}\quad
D_1 = -2k\cdot v-i0\,,\quad
D_2 = -k^2-i0\,,
\label{M2:F}
\end{equation}
and $f(k^2)$ is an arbitrary function.
We can construct an identity in which $f'(k^2)$ terms cancel:
\begin{equation}
\frac{\partial}{\partial k} \cdot
\left( k - 2 \frac{D_2}{D_1} v \right)
\frac{f(k^2)}{D_1^{n_1} D_2^{n_2}}
= \left[ d-n_1-2 - 4 (n_1+1) \frac{D_2}{D_1^2} \right]
\frac{f(k^2)}{D_1^{n_1} D_2^{n_2}}\,.
\label{M2:id}
\end{equation}
Integrating it, we obtain an integration-by-parts relation~\cite{GSS:06}
\begin{equation}
(d-n_1-2) F(n_1,n_2) = 4 (n_1+1) \1{++} \2- F(n_1,n_2)\,.
\label{M2:rel}
\end{equation}

\begin{figure}[h]
\begin{center}
\begin{picture}(92,52)
\put(46,26){\makebox(0,0){\includegraphics{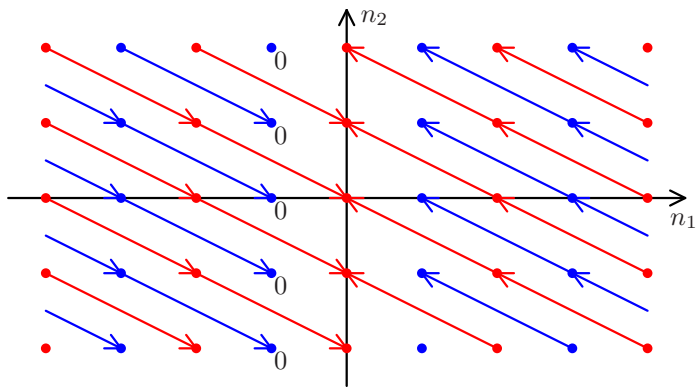}}}
\put(91,24){\makebox(0,0)[t]{$n_1$}}
\put(48,50){\makebox(0,0)[l]{$n_2$}}
\put(36.5,45.5){\makebox(0,0)[tl]{$0$}}
\put(36.5,35.5){\makebox(0,0)[tl]{$0$}}
\put(36.5,25.5){\makebox(0,0)[tl]{$0$}}
\put(36.5,15.5){\makebox(0,0)[tl]{$0$}}
\put(36.5,5.5){\makebox(0,0)[tl]{$0$}}
\end{picture}
\end{center}
\caption{Recurrence relation}
\label{F:aver}
\end{figure}

Let's call integrals with even $n_1$ apparently even,
and with odd $n_1$ --- apparently odd
(they would be even and odd in $v$ if we neglected $i0$
in the denominator).
These two classes of integrals are not mixed
by the recurrence relation~(\ref{M2:rel}).
We can use this relation to reduce all apparently even integrals
to vacuum integrals with $n_1=0$ (Fig.~\ref{F:aver}).
Apparently odd integrals with $n_1<0$ can be reduced to $n_1=-1$.
Substituting $n_1=-1$ to~(\ref{M2:rel}),
we see that these integrals vanish,
and hence all integrals with odd $n_1<0$ vanish too.
Apparently odd integrals with $n_1>0$ can be reduced to $n_1=1$
(Fig.~\ref{F:aver});
however, they are not related to those with $n_1=-1$.

The solution of the recurrence relation can thus be written as
\begin{equation}
F(n_1,n_2) =
\left\{
\begin{array}{ll}
\displaystyle (-4)^{-n_1/2}
\frac{\Gamma\left(\frac{d}{2}\right)}{\Gamma\bigl(\frac{d-n_1}{2}\bigr)}
\frac{\Gamma\left(\frac{1-n_1}{2}\right)}{\Gamma\left(\frac{1}{2}\right)}
F\Bigl(0,n_2+\frac{n_1}{2}\Bigr)
& \mbox{even $n_1$,}\\
\displaystyle 2^{1-n_1}
\frac{\Gamma\left(\frac{d-1}{2}\right)}%
{\Gamma\left(\frac{n_1+1}{2}\right)\Gamma\bigl(\frac{d-n_1}{2}\bigr)}
F\Bigl(1,n_2+\frac{n_1-1}{2}\Bigr)
& \mbox{odd $n_1>0$,}\\
\displaystyle 0 & \mbox{odd $n_1<0$.}
\end{array}
\right.
\label{M2:sol}
\end{equation}

Some of these properties can be understood more directly.
If $n_1<0$, $i0$ in $D_1^{-n_1}$ can be safely neglected;
averaging this factor over $k$ directions, we obtain $0$ for odd $n_1$
and the upper formula in~(\ref{M2:sol}) for even $n_1$.
It was suggested~\cite{BG:95} that this last formula can also be used
for even $n_1>0$, but the proof (presented here)
only appeared in~\cite{GSS:06}.

\begin{figure}[ht]
\begin{center}
\begin{picture}(52,26)
\put(26,14.5){\makebox(0,0){\includegraphics{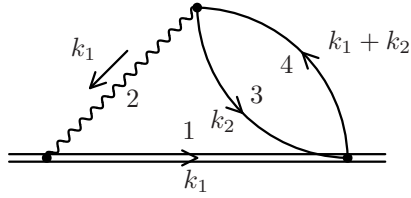}}}
\put(26,3){\makebox(0,0)[t]{$k_1$}}
\put(11,19){\makebox(0,0){$k_1$}}
\put(29.5,9.5){\makebox(0,0){$k_2$}}
\put(49,20){\makebox(0,0){$k_1+k_2$}}
\put(25,7){\makebox(0,0)[b]{1}}
\put(17.5,12.5){\makebox(0,0){2}}
\put(34,13){\makebox(0,0){3}}
\put(38,17){\makebox(0,0){4}}
\end{picture}
\end{center}
\caption{Two-loop diagram}
\label{F:M2}
\end{figure}

Now let's consider the two-loop diagram (Fig.~\ref{F:M2})
\begin{equation}
F(n_1,n_2,n_3,n_4) = \frac{1}{(i\pi^{d/2})^2} \int
\frac{d^d k_1\,d^d k_2}{D_1^{n_1} D_2^{n_2} D_3^{n_3} D_4^{n_4}}\,,
\label{M2:def}
\end{equation}
where
\begin{align*}
&D_1 = -2 k_1\cdot v - i0\,,\quad
D_2 = -k_1^2 - i0\,,\\
&D_3 = 1 - k_2^2 - i0\,,\quad
D_4 = 1 - (k_1+k_2)^2 - i0\,.
\end{align*}
It is symmetric with respect to $3\leftrightarrow4$,
and vanishes if $n_3$ or $n_4$ is integer and non-positive.
It can be calculated using $\alpha$ parametrization~\cite{GSS:06}:
\begin{align}
&F(n_1,n_2,n_3,n_4) ={}
\label{M2:F2}\\
&\frac{\Gamma\bigl(\frac{n_1}{2}\bigr)
\Gamma\bigl(\frac{d-n_1}{2}-n_2\bigr)
\Gamma\bigl(\frac{n_1-d}{2}+n_2+n_3\bigr)
\Gamma\bigl(\frac{n_1-d}{2}+n_2+n_4\bigr)
\Gamma\bigl(\frac{n_1}{2}+n_2+n_3+n_4-d\bigr)}%
{2\Gamma(n_1) \Gamma(n_3) \Gamma(n_4)
\Gamma\bigl(\frac{d-n_1}{2}\bigr)
\Gamma(n_1+2n_2+n_3+n_4-d)}\,.
\nonumber
\end{align}
In full accordance with~(\ref{M2:sol}),
integrals $F(n_1,n_2,n_3,n_4)$ with even $n_1$
reduce to $F(0,n_2+n_1/2,n_3,n_4)$
(this is a well-known two-loop vacuum integral~\cite{V:80});
those with odd $n_1>0$ reduce to $F(1,n_2+(n_1-1)/2,n_3,n_4)$;
and those with odd $n_1<0$ vanish.
All apparently even integrals are proportional
to the single master integral
\begin{equation}
I_0^2 = \raisebox{-9.8mm}{\includegraphics{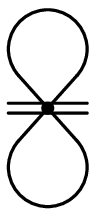}}\,,
\label{M2:I02}
\end{equation}
and apparently odd ones --- to
\begin{equation}
J_0 = \raisebox{-5.8mm}{\includegraphics{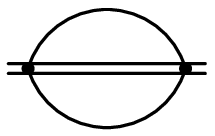}}
= 2^{4d-9} \pi^2 \frac{\Gamma(5-2d)}{\Gamma^2(2-d/2)}\,.
\label{M2:J0}
\end{equation}
Integrals~(\ref{M2:def}) can also contain
powers of $(2 k_2 + k_1)\cdot v$ in the numerator;
see~\cite{GSS:06} for details of their evaluation.

\subsection{Three loops}
\label{S:M3}

\subsubsection{Reduction}
\label{S:M3r}

\begin{figure}[ht]
\begin{center}
\begin{picture}(61,12)
\put(14,6){\makebox(0,0){\includegraphics{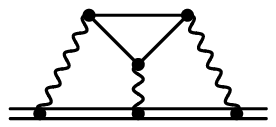}}}
\put(47,6){\makebox(0,0){\includegraphics{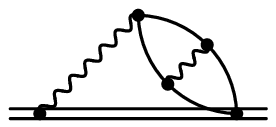}}}
\end{picture}
\end{center}
\caption{Topologies of three-loop on-shell HQET propagator diagrams with mass}
\label{F:M3t}
\end{figure}

There are two generic topologies of three-loop
on-shell HQET propagator diagrams with a massive loop
(Fig.~\ref{F:M3t}).
Algorithms of their reduction to master integrals,
using integration by parts identities,
have been constructed~\cite{GSS:06}
by Gr\"obner bases technique~\cite{SS:06}.

All apparently even integrals of the first topology
reduce to
\begin{equation*}
\raisebox{-6.3mm}{\includegraphics{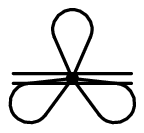}}\quad
\raisebox{-6.3mm}{\includegraphics{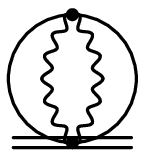}}\quad
\raisebox{-4.3mm}{\includegraphics{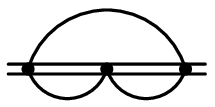}}\quad
\raisebox{-3.8mm}{\includegraphics{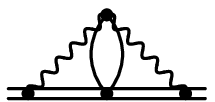}}\quad
\raisebox{-3.8mm}{\includegraphics{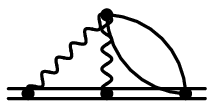}}
\end{equation*}
while apparently odd ones to
\begin{equation*}
\raisebox{-6.3mm}{\includegraphics{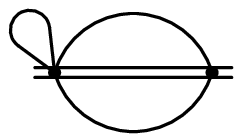}}\quad
\raisebox{-5.8mm}{\includegraphics{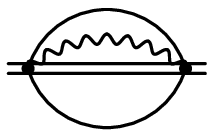}}\quad
\raisebox{-3.8mm}{\includegraphics{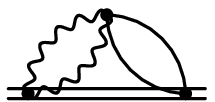}}
\end{equation*}
All apparently even integrals of the second topology
reduce to
\begin{equation*}
\raisebox{-6.3mm}{\includegraphics{grozin_andrey.fig78.eps}}\quad
\raisebox{-6.3mm}{\includegraphics{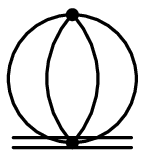}}
\end{equation*}
while apparently odd ones to
\begin{equation*}
\raisebox{-6.3mm}{\includegraphics{grozin_andrey.fig83.eps}}\quad
\raisebox{-5.8mm}{\includegraphics{grozin_andrey.fig84.eps}}\quad
\raisebox{-3.8mm}{\includegraphics{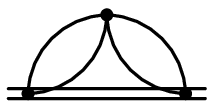}}
\end{equation*}

The master integrals
\begin{equation*}
\raisebox{-6.3mm}{\includegraphics{grozin_andrey.fig79.eps}}\quad
\raisebox{-5.8mm}{\includegraphics{grozin_andrey.fig84.eps}}\quad
\raisebox{-5.8mm}{\includegraphics{grozin_andrey.fig85.eps}}
\end{equation*}
can be easily expressed via $\Gamma$ functions.
The master integral
\begin{equation*}
\raisebox{-6.3mm}{\includegraphics{grozin_andrey.fig86.eps}}
\end{equation*}
has been investigated in detail~\cite{B:92,B:96}.

\subsubsection{A master integral}
\label{S:M31}

Now we shall discuss the integrals
\begin{equation}
I_{n_1 n_2 n_3} =
\raisebox{-10mm}{\begin{picture}(27,17.25)
\put(13.5,8.625){\makebox(0,0){\includegraphics{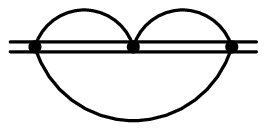}}}
\put(13.5,0){\makebox(0,0)[b]{$n_3$}}
\put(8.5,17.25){\makebox(0,0)[t]{$n_2$}}
\put(18.5,17.25){\makebox(0,0)[t]{$n_2$}}
\put(8.5,9.5){\makebox(0,0)[t]{$n_1$}}
\put(18.5,9.5){\makebox(0,0)[t]{$n_1$}}
\end{picture}}
\label{M3:In1n2n3}
\end{equation}
($I_{111}$ is one of the master integrals).
Several approaches have been tried~\cite{GSS:06,GHM:07}.
The best result was obtained~\cite{GHM:07}
using a method similar to~\cite{B:92}.

\begin{figure}[ht]
\begin{center}
\begin{picture}(42,42)
\put(21,21){\makebox(0,0){\includegraphics{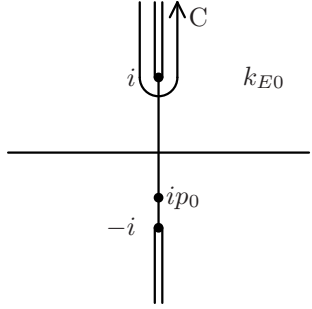}}}
\put(22,15){\makebox(0,0)[l]{$i p_0$}}
\put(18,31){\makebox(0,0)[r]{$i$}}
\put(18,11){\makebox(0,0)[r]{$-i$}}
\put(25,39){\makebox(0,0)[l]{C}}
\put(35,31){\makebox(0,0){$k_{E0}$}}
\end{picture}
\end{center}
\caption{Integration contour}
\label{F:C}
\end{figure}

First we consider (following~\cite{G:08}) the one-loop subdiagram
\begin{equation}
I_{n_1 n_2}(p_0) =
\raisebox{-2.5mm}{\begin{picture}(30,14)
\put(15,7){\makebox(0,0){\includegraphics{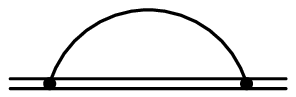}}}
\put(15,0){\makebox(0,0)[b]{$n_1$}}
\put(15,14){\makebox(0,0)[t]{$n_2$}}
\end{picture}} =
\frac{1}{i\pi^{d/2}} \int
\frac{d k_0\,d^{d-1}\vec{k}}{[-2(k_0+p_0)-i0]^{n_1} [1-k^2-i0]^{n_2}}\,.
\label{M3:In1n2}
\end{equation}
After the Wick rotation, we integrate in $d^{d-1}\vec{k}$:
\begin{equation*}
I_{n_1 n_2}(p_0) =
\frac{\Gamma(n_2-(d-1)/2)}{\pi^{1/2} \Gamma(n_2)}
\int_{-\infty}^{+\infty} d k_{E0}
\frac{(k_{E0}^2+1)^{(d-1)/2-n_2}}{(- 2 p_0 - 2 i k_{E0})^{n_1}}\,.
\end{equation*}
If $p_0<0$, we can deform the integration contour (Fig.~\ref{F:C}):
\begin{equation*}
I_{n_1 n_2}(p_0) = 2 \frac{\Gamma(n_2-(d-1)/2)}{\pi^{1/2} \Gamma(n_2)}
\cos\left[\pi \left(\frac{d}{2} - n_2\right)\right]
\int_1^\infty dk
\frac{(k^2-1)^{(d-1)/2-n_2}}{(2k - 2p_0)^{n_1}}\,.
\end{equation*}
This integral is
\begin{align*}
I_{n_1 n_2}(p_0) =&
\frac{\Gamma(n_1+n_2-2+\varepsilon) \Gamma(n_1+2n_2-4+2\varepsilon)}%
{\Gamma(n_2) \Gamma(2(n_1+n_2-2+\varepsilon))}\\
&\times{}_2 F_1 \left( \left.
\begin{array}{c}
n_1,n_1+2n_2-4+2\varepsilon\\
n_1+n_2-\frac{3}{2}+\varepsilon
\end{array}
\right| \frac{1}{2} \left(1 + p_0\right) \right)\,,
\end{align*}
or, after using a ${}_2 F_1$ identity,
\begin{align}
I_{n_1 n_2}(p_0) =&
\frac{\Gamma(n_1+n_2-2+\varepsilon) \Gamma(n_1+2n_2-4+2\varepsilon)}%
{\Gamma(n_2) \Gamma(2(n_1+n_2-2+\varepsilon))}
\nonumber\\
&\times{}_2 F_1 \left( \left.
\begin{array}{c}
\frac{1}{2} n_1, \frac{1}{2} n_1 + n_2 - 2 + \varepsilon\\
n_1 + n_2 - \frac{3}{2} + \varepsilon
\end{array}
\right| 1 - p_0^2 \right)\,.
\label{M3:res1}
\end{align}

This result was obtained~\cite{GHM:07}
using the HQET Feynman parametrization:
\begin{equation*}
I_{n_1 n_2}(p_0) =
\frac{\Gamma(n_1+n_2-2+\varepsilon)}{\Gamma(n_1) \Gamma(n_2)}
\int_0^\infty y^{n_1-1} (y^2 - 2 p_0 y + 1)^{2-n_1-n_2-\varepsilon}\,dy\,.
\end{equation*}
This integral at $p_0<0$ gives~(\ref{M3:res1})
(a similar expression has been derived in~\cite{Z:02}).

Now we can integrate in $d^{d-1}\vec{p}$
in the three-loop diagram:
\begin{equation}
I_{n_1 n_2 n_3} =
\frac{\Gamma(n_3-3/2+\varepsilon)}{\pi^{1/2}\Gamma(n_3)}
\int_{-\infty}^{+\infty}
I_{n_1 n_2}^2(i p_{E0}) (1+p_{E0}^2)^{3/2-n_3-\varepsilon} d p_{E0}\,.
\label{M3:I}
\end{equation}
The square of ${}_2 F_1$ in~(\ref{M3:res1})
can be expressed via an ${}_3 F_2$
using the Clausen identity.
We analytically continue this ${}_3 F_2$ from $1+p_{E0}^2>1$
to $z=1/(1+p_{E0}^2)<1$ and integrate~(\ref{M3:I}) term by term.
The result contains, in general, three ${}_4 F_3$
of unit argument.

A convergent integral $I_{122}$ is related
to the master integral $I_{111}$ by
\begin{equation*}
I_{122} = -
\frac{(d-3)^2 (d-4) (3d-8) (3d-10)}{8 (3d-11) (3d-13)} I_{111}\,.
\end{equation*}
For this integral, we obtain~\cite{GHM:07}
\begin{align}
\frac{I_{122}}{\Gamma^3(1+\varepsilon)} =
- \frac{1}{2\varepsilon^2} \Biggl[& \frac{1}{1+2\varepsilon}
{}_4 F_3 \left( \left.
\begin{array}{c}
1, \frac{1}{2}-\varepsilon, 1+\varepsilon, -2\varepsilon\\
\frac{3}{2}+\varepsilon, 1-\varepsilon, 1-2\varepsilon
\end{array}
\right| 1 \right)
\nonumber\\
&{} - \frac{2}{1+4\varepsilon}
\frac{\Gamma^2(1-\varepsilon) \Gamma^3(1+2\varepsilon)}%
{\Gamma^2(1+\varepsilon) \Gamma(1-2\varepsilon) \Gamma(1+4\varepsilon)}
{}_3 F_2 \left( \left.
\begin{array}{c}
\frac{1}{2}, 1+2\varepsilon, -\varepsilon\\
\frac{3}{2}+2\varepsilon, 1-\varepsilon
\end{array}
\right| 1 \right)
\nonumber\\
&{} + \frac{1}{1+6\varepsilon}
\frac{\Gamma^2(1-\varepsilon) \Gamma^4(1+2\varepsilon)
\Gamma(1-2\varepsilon) \Gamma^2(1+3\varepsilon)}%
{\Gamma^4(1+\varepsilon) \Gamma(1+4\varepsilon)
\Gamma(1-4\varepsilon) \Gamma(1+6\varepsilon)}
\Biggr]\,.
\label{M3:hyper}
\end{align}
Expansion of this result up to $\varepsilon^7$
agrees with~\cite{GHM:07}
\begin{equation}
\frac{I_{122}}{\Gamma^3(1+\varepsilon)} =
\frac{\pi^2}{3}
\frac{\Gamma^3(1+2\varepsilon) \Gamma^2(1+3\varepsilon)}%
{\Gamma^6(1+\varepsilon) \Gamma(2+6\varepsilon)}\,.
\label{M3:Maitre}
\end{equation}
This equality has also been checked by high precision
numerical calculations at some finite $\varepsilon$ values.
This conjectured hypergeometric identity can also be rewritten
in a nice form~\cite{G:08}
\begin{equation}
g_1(\varepsilon)
{}_4 F_3 \left( \left.
\begin{array}{c}
1, \frac{1}{2}-\varepsilon, 1+\varepsilon, -2\varepsilon\\
\frac{3}{2}+\varepsilon, 1-\varepsilon, 1-2\varepsilon
\end{array}
\right| 1 \right)
- 2 g_2(\varepsilon)
{}_3 F_2 \left( \left.
\begin{array}{c}
\frac{1}{2}, 1+2\varepsilon, -\varepsilon\\
\frac{3}{2}+2\varepsilon, 1-\varepsilon
\end{array}
\right| 1 \right)
+ g_3(\varepsilon) = 0\,,
\label{M3:David}
\end{equation}
where
\begin{equation*}
b(\varepsilon) =
\frac{\Gamma(1-\varepsilon)\Gamma(1+2\varepsilon)}{\Gamma(1+\varepsilon)}\,,
\quad
g_n(\varepsilon) =
\frac{b^n(\varepsilon)}{b(n\varepsilon)(1+2n\varepsilon)}\,.
\end{equation*}
We have no analytical proof.

\subsubsection{Other master integrals}
\label{S:M3b}

Other master integrals were calculated~\cite{GSS:06}
using Mellin--Barnes representation (see, e.g., \cite{S:06}).
Now we shall discuss a simple example of this technique.
Let's consider the one-loop propagator diagram
with two massive lines:
\begin{equation*}
\frac{1}{i\pi^{d/2}}
\int \frac{d^d k}{[m^2-k^2]^{n_1} [m^2-(k+p)^2]^{n_2}}\,.
\end{equation*}
Using Feynman parametrization,
\begin{align*}
&= \frac{\Gamma(n_1+n_2)}{\Gamma(n_1) \Gamma(n_2)} \frac{1}{i\pi^{d/2}}
\int \frac{dx\,x^{n_2-1}(1-x)^{n_1-1}\,d^d k}%
{[(m^2-k^2)(1-x)+(m^2-(k+p)^2)x]^{n_1+n_2}}\\
&= \frac{\Gamma(n_1+n_2)}{\Gamma(n_1) \Gamma(n_2)} \frac{1}{i\pi^{d/2}}
\int \frac{dx\,x^{n_2-1}(1-x)^{n_1-1}\,d^d k}%
{[-k^2-2xp\cdot k-x p^2+m^2]^{n_1+n_2}}\,.
\end{align*}
After the shift $k=k'-xp$:
\begin{equation*}
= \frac{\Gamma(n_1+n_2)}{\Gamma(n_1) \Gamma(n_2)} \frac{1}{i\pi^{d/2}}
\int \frac{dx\,x^{n_2-1}(1-x)^{n_1-1}\,d^d k}%
{[m^2+x(1-x)(-p^2)-k^{\prime2}]^{n_1+n_2}}\,,
\end{equation*}
we can integrate in $k'$:
\begin{equation*}
= \frac{\Gamma\bigl(n_1+n_2-\frac{d}{2}\bigr)}{\Gamma(n_1) \Gamma(n_2)}
\int_0^1 \frac{dx\,x^{n_2-1}(1-x)^{n_1-1}}{[m^2+x(1-x)(-p^2)]^{n_1+n_2-d/2}}\,.
\end{equation*}

Now we shall use Mellin--Barnes representation
\begin{equation}
\frac{1}{(a+b)^n} = \frac{a^{-n}}{\Gamma(n)} \frac{1}{2\pi i}
\int_{-i\infty}^{+i\infty} dz\,\Gamma(-z) \Gamma(n+z)
\left(\frac{b}{a}\right)^z\,.
\label{M3:MB}
\end{equation}
Here the integration contour is chosen in such a way
that all poles of $\Gamma(\cdots+z)$
(they are called \textit{left poles})
are to the left of the contour,
and all poles of $\Gamma(\cdots-z)$
(they are called \textit{right poles})
are to the right of it.
It is easy to check~(\ref{M3:MB}):
closing the contour to the right we get the expansion
of the left-hand side in $b/a$;
closing it to the left --- the expansion in $a/b$.

We continue our calculation:
\begin{align*}
&= \frac{m^{d-2(n_1+n_2)}}{\Gamma(n_1) \Gamma(n_2)} \frac{1}{2\pi i}
\int_{-i\infty}^{+i\infty} dz \Gamma(-z) \Gamma(n_1+n_2+z)
\left(\frac{-p^2}{m^2}\right)^z
\int_0^1 dx\,x^{n_2+z-1} (1-x)^{n_1+z-1}\\
&= \frac{m^{d-2(n_1+n_2)}}{\Gamma(n_1) \Gamma(n_2)} \frac{1}{2\pi i}
\int_{-i\infty}^{+i\infty} dz
\frac{\Gamma(-z) \Gamma(n_1+z) \Gamma(n_2+z)
\Gamma\bigl(n_1+n_2-\frac{d}{2}+z\bigr)}{\Gamma(n_1+n_2+2z)}
\left(\frac{-p^2}{m^2}\right)^z\,.
\end{align*}
This means that two massive lines can be replaced by one massless one
(raised to the power $-z$)
at the price of one extra integration in $z$:
\begin{align}
&\raisebox{-8mm}{\begin{picture}(22,18)
\put(11,9){\makebox(0,0){\includegraphics{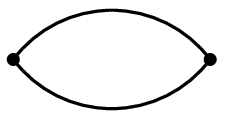}}}
\put(11,16){\makebox(0,0){$n_1$}}
\put(11,2){\makebox(0,0){$n_2$}}
\end{picture}}
= \frac{1}{i\pi^{d/2}} \int
\frac{d^d k}{\left[m^2-k^2-i0\right]^{n_1}
\left[m^2-(k+p)^2-i0\right]^{n_2}} =
\nonumber\\
&\frac{m^{d-2(n_1+n_2)}}{\Gamma(n_1) \Gamma(n_2)}
\frac{1}{2\pi i} \int_{-i\infty}^{+i\infty} dz
\frac{\Gamma(-z) \Gamma(n_1+z) \Gamma(n_2+z)
\Gamma(n_1+n_2-d/2+z)}{\Gamma(n_1+n_2+2z)}
\nonumber\\
&\qquad\qquad\qquad\qquad
m^{-2z}
\raisebox{-3mm}{\begin{picture}(22,8)
\put(11,4){\makebox(0,0){\includegraphics{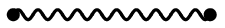}}}
\put(11,6){\makebox(0,0){$-z$}}
\end{picture}}
\label{M3:MB1}
\end{align}

This trick allows us to reduce this master integral
to a single Mellin--Barnes integral.
Using integration by parts, we can kill one of three lines
in the left (integer) triangle,
and calculate the integrand in $\Gamma$ functions.
This allows us to find several terms of its $\varepsilon$ expansion:
\begin{align}
&\raisebox{-3.8mm}{\includegraphics{grozin_andrey.fig82.eps}} =
\frac{\Gamma^2(2\varepsilon)\Gamma(3\varepsilon-1)}{4\Gamma(4\varepsilon)}
\frac{1}{2\pi i} \int_{-i\infty}^{+i\infty} dz
\nonumber\\
&\frac{\Gamma(1+z) \Gamma(1/2+\varepsilon+z) \Gamma(1+\varepsilon+z)
\Gamma(-2\varepsilon-z) \Gamma(-\varepsilon-z) \Gamma(-z)}%
{\Gamma(3/2+\varepsilon+z) \Gamma(1-2\varepsilon-z)}
\nonumber\\
&{}= - \Gamma^3(1+\varepsilon) \Biggl[
\frac{\pi^2}{9\varepsilon^2}
- \frac{6\zeta_3-5\pi^2}{9\varepsilon}
+ \frac{11}{270} \pi^4 - \frac{10}{3} \zeta_3 + \frac{19}{9} \pi^2
\nonumber\\
&{} + \left( - \frac{8}{3} \zeta_5 + \frac{8}{9} \pi^2 \zeta_3
+ \frac{11}{54} \pi^4 - \frac{38}{3} \zeta_3 + \frac{65}{9} \pi^2
\right) \varepsilon + \cdots \Biggr]\,.
\label{M3:I4}
\end{align}

This master integral has been evaluated in a closed form
using Mellin--Barnes in $\alpha$ representation:
\begin{align}
\raisebox{-3.8mm}{\includegraphics{grozin_andrey.fig81.eps}} =&
\frac{\Gamma(1/2-\varepsilon)
\Gamma(-\varepsilon)
\Gamma^2(2\varepsilon)
\Gamma(1+\varepsilon)
\Gamma(3\varepsilon-1)}%
{4 \Gamma(3/2-\varepsilon) \Gamma(4\varepsilon)}
\nonumber\\
&{}\times
\left[ \psi\left(\frac{1}{2}-\varepsilon\right)
+ \psi\left(1-\varepsilon\right) - 2 \log 2
+2 \gamma_E \right]\,.
\label{M3:I3}
\end{align}

This master integral can be written as a double Mellin--Barnes integral
using~(\ref{M3:MB1}).
It appears possible to calculate one integral:
\begin{align}
&\raisebox{-3.8mm}{\includegraphics{grozin_andrey.fig87.eps}} =
\frac{\pi^{3/2} }{4^\varepsilon \Gamma(3/2-\varepsilon)}\frac{1}{2\pi i}
\nonumber\\
&\int_{-i\infty}^{+i\infty} dz
\frac{\Gamma(1+z) \Gamma\left(\frac{3}{2}-\varepsilon+z\right)
\Gamma(\varepsilon+z)
\Gamma\left(-\frac{1}{2}+\varepsilon-z\right)
\Gamma\left(-\frac{3}{2}+2\varepsilon-z\right) \Gamma(-z)}%
{\Gamma\left(\frac{3}{2}+z\right) \Gamma(\varepsilon-z)}
\nonumber\\
&{} = \Gamma^3(1+\varepsilon) \frac{32}{3} \pi^2
\left[ -1 + 2 \left( 4 \log 2 - \pi - 7 \right) \varepsilon
+ \cdots \right]\,.
\label{M3:J3}
\end{align}

\subsubsection{Applications}
\label{S:app2}

Feynman integrals considered here were used~\cite{GSS:06}
for calculating the matching coefficients
for the HQET heavy-quark field and the heavy--light quark current
between the $b$-quark HQET with dynamic $c$-quark loops
and without such loops
(the later theory is the low-energy approximation
for the former one at scales below $m_c$).
Another recent application --- the effect of $m_c\ne0$
on $b\to c$ plus lepton pair at three loops~\cite{CP:08}.
The method of regions was used;
the purely soft region (loop momenta $\sim m_c$)
gives integrals of this type.
Two extra terms of $\varepsilon$ expansion
of the master integral of Sect.~\ref{S:M31}
were required for this calculation
which were not obtained in~\cite{GSS:06}.
This was the initial motivation for~\cite{GHM:07}.

\section*{Acknowledgements}

I am grateful to
D.J.~Broadhurst,
K.G.~Chetyrkin,
A.I.~Davydychev,
T.~Huber,
D.~Ma\^{\i}tre,
A.V.~Smirnov,
V.A.~Smirnov
for collaboration on HQET projects discussed here;
to D.J.~Broadhurst for discussions of the master integral
of Sect.~\ref{S:M31},
in particular for rewriting the conjectured identity
in the nice form~(\ref{M3:David});
to R.N.~Lee for the idea of the new derivation of~(\ref{J:J});
to S.V.~Mikhailov and V.A.~Smirnov for discussions
about $\alpha$ parametrization in HQET (Sect.~\ref{S:alpha});
and to the organizers of the summer school
on heavy quark physics in Dubna for inviting me
to give these lectures.

\begin{footnotesize}

\end{footnotesize}

\end{document}